\documentclass[aps,reprint,superscriptaddress,flushbottom,pra,floatfix]{revtex4-1}
\usepackage[USenglish]{babel} 
\usepackage[T1]{fontenc}

\usepackage{amsmath}    
\usepackage{amssymb}	
\usepackage{relsize}
\usepackage{graphicx}   
\usepackage{verbatim}   
\usepackage{color}      
\usepackage{subfigure}  
\usepackage{hyperref}   
\usepackage{soul}
\usepackage{natbib}
\usepackage{setspace}
\usepackage{placeins}
\usepackage{mathrsfs}
\usepackage{endnotes}
\usepackage[bottom]{footmisc}

\newcommand{\nt}[1]{{#1}}

\newcommand{\ntt}[1]{{#1}}

\let\footnote=\endnote

\definecolor{light-blue}{rgb}{0.859375,0.898438,0.945317}
\definecolor{light-red}{rgb}{0.945,0.859,0.8554}
\definecolor{bright-blue}{rgb}{0,0.6875,1}
\definecolor{purple}{rgb}{.86,.4,.95}

\begin{document}

\title{Sensitivity and accuracy of Casimir force measurements in air}
\author{Joseph L. Garrett}
	\affiliation{ University of Maryland Department of Physics, College Park, MD 20742, USA}
	\affiliation{ Institute for Research in Electronics and Applied Physics, College Park, MD 20742}

\author{David A. T. Somers}
	\affiliation{ University of Maryland Department of Physics, College Park, MD 20742, USA}
	\affiliation{ Institute for Research in Electronics and Applied Physics, College Park, MD 20742}

\author{Kyle Sendgikoski}
	\affiliation{ University of Maryland Department of Physics, College Park, MD 20742, USA}
	\affiliation{ Institute for Research in Electronics and Applied Physics, College Park, MD 20742}

\author{Jeremy N. Munday}	
	\email{jnmunday@umd.edu}
	\affiliation{ Institute for Research in Electronics and Applied Physics, College Park, MD 20742}
	\affiliation{ Department of Electrical and Computer Engineering, College Park, MD 20742}

\begin{abstract}
Quantum electrodynamic fluctuations cause an attractive force between metallic surfaces. 
At separations where the finite speed of light affects the interaction, it is called the Casimir force.
Thermal motion determines the fundamental sensitivity limits of its measurement at room temperature, but several other systematic errors contribute uncertainty as well and become more significant in air relative to vacuum.
Here we discuss the viability of \nt{the force modulation} measurement technique in air \nt{(compared to frequency modulation, which is typically used in vacuum, and quasi-static deflection, which is usually used in fluid)}, characterize its sensitivity and accuracy by identifying several dominant sources of uncertainty, and compare the results to the fundamental sensitivity limits dictated by thermal motion and to the uncertainty inherent to calculations of the Casimir force.
Finally, we explore prospects for mitigating the sources of uncertainty to enhance the range and accuracy of Casimir force measurements.
\end{abstract}

\maketitle
\section{Introduction}

H. B. G. Casimir predicted a force between surfaces that originates from quantum electromagnetic fluctuations \cite{Casimir1948a}.
This force is of theoretical and practical interest because it is an application of quantum electrodynamics to bulk materials. 
Lifshitz extended the analysis to arbitrary materials \cite{Lifshitz1956,Dzyaloshinskii1961}, including the prediction of a repulsive Casimir force\nt{,} which has since been experimentally confirmed \cite{Munday2009a}. 
The force has been measured numerous times \cite{Sparnaay1958,VanBlokland1978,Lamoreaux1997,Mohideen1998,Ederth2000,Chan2001,Lisanti2005a,Masuda2009,Laurent2012}, between many materials \cite{Decca2003,Chen2006,DeMan2009a,Torricelli2010,Torricelli2011b,Bimonte2016a,Norte2018,Somers2018}, in several geometries \cite{Bressi2002,Intravaia2013a,Tang2017,Garrett2018}, and with increasing precision \cite{Decca2005,Jourdan2009,Chang2012,Sushkov2011a}. 

\nt{Because measurements in gas provide a middle ground between the high sensitivities of measurements in vacuum and the exotic Casimir force behavior in liquid environments, they have frequently contributed to critical experimental tests of the Casimir force \cite{VanZwol2008c,Sedighi2016,Cunuder2018,DeMan2010,Stange2018}. 
For example, de Man {\it et al.} \cite{DeMan2009a} verified that a significant difference in the visible dielectric function can halve the magnitude of the Casimir force, and Van Zwol {\it et al.} tested how roughness affects the Casimir force \cite{VanZwol2007}.
Moreover, exploring the contributions to uncertainty in one environment can clarify how they appear in another \cite{Cunuder2018}.
Understanding the uncertainty in air would assist the interpretation of Casimir force measurements in a variable pressure chamber, which are being undertaken to separate hypothetical local-density-coupled chameleon forces from the Casimir force \cite{Almasi2015a,Sedmik2018}.
Some sources of uncertainty, such as patch potentials, are predicted to have reduced magnitude in air relative to vacuum \cite{Behunin2012}.}

Efforts to harness the Casimir force for new MEMS devices \cite{Capasso2007} have resulted in non-linear MEMS oscillators \cite{Serry1995,Chan2001a} and on-chip Casimir force measurement devices \cite{Zou2013,Tang2017}. 
Several measurements of the Casimir force have been made in ambient conditions, a necessary test for realistic MEMS.
\nt{Drag in air can be utilized to develop MEMS techniques that are difficult in vacuum. 
For example, it has recently been shown that an oscillating spherical Casimir probe in air can scan the topography of a surface, which can be useful for aligning novel geometries \cite{Garrett2018}. 
The drag helps to avoid collisions because it slows the probe as it approaches the surface.
Likewise,} the measurements of de Man {\it et al.}\ present a framework for using the Casimir force to actuate dynamical MEMS \cite{DeMan2010,Deman2010a}.

\begin{figure}[b]
	\centering
	\includegraphics[width=.42\textwidth]{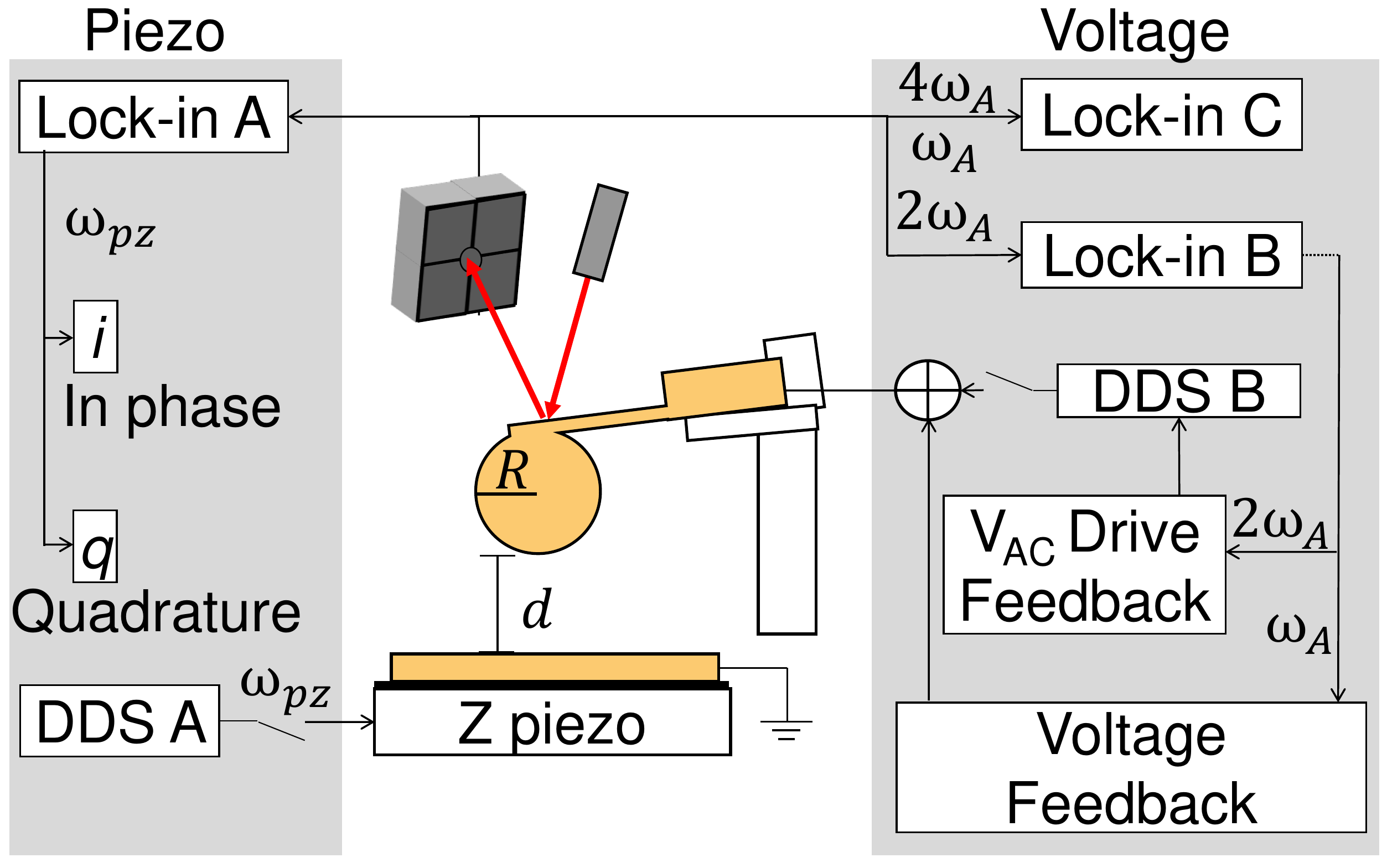}
	\caption{ 
		An atomic force microscope is used to measure the Casimir force. An optical lever detects deflections of the cantilever to which the sphere is attached. Direct digital synthesizer (DDS) A drives the piezoelectric transducer to shake the gold-coated plate. A lock-in amplifier (LIA) detects the cantilever's response and separates it into in-phase and quadrature components. Each LIA in the AFM detects up to two signals. DDS B applies an AC voltage to the cantilever at frequency $\omega_{\text{A}}$. The oscillations of the cantilever are then detected at frequencies $\omega_{\text{A}}$ and $2\omega_{\text{A}}$ with LIA B and $4\omega_{\text{A}}$ with LIA C. A feedback loop adjusts $V_{\text{AC}}$ so that the oscillation at $2\omega_{\text{A}}$ is constant during the measurement of the Casimir force. The signal at $\omega_{\text{A}}$ is used to estimate and mitigate the minimizing voltage $V_{0}$ by applying a DC voltage to the sphere.
	}
	\label{fig:AFM_setup}
\end{figure}

Here we test the \nt{sensitivity and accuracy of the force modulation (FoM) measurement technique} in air using an atomic force microscope (AFM) --- depicted in Fig. \ref{fig:AFM_setup}, identify several sources of uncertainty, estimate the uncertainty from each source, and discuss strategies to reduce uncertainty in future measurements. 
Our intention is to identify sources of uncertainty, \nt{their relative contributions to the total error,} and tactics to mitigate them, in order to facilitate the development of new experiments.
In addition, tabulating the uncertainty in a measurement helps to clarify when it can be used to distinguish between different hypothesis regarding the computation of the Casimir force.

As stated in \cite{DeMan2010}, a measurement technique must satisfy three requirements in addition to detecting the Casimir force: (i) it must mitigate the contributions of other forces (hydrodynamic, electrostatic, {\it etc.}), (ii) it must determine the \nt{absolute} separation\nt{, $d_{0}$,} between the sphere and plate, and (iii) it must calibrate the force signal. 
We characterize how well (i)-(iii) are achieved and quantify the amount of uncertainty each imparts to a measurement. 
Furthermore, several instrumental sources of error, such as optical interference, may manifest themselves differently in different experiments, but are common to many force measurement techniques. 
A few sources of error that impart uncertainty to the total calculated force rather than the force measurement, such as roughness, patch potentials, and limited dielectric information, have been discussed extensively in the literature \cite{Pirozhenko2006,Munday2008c,Svetovoy2008,Genet2003a,Broer2012,Sedmik2013,Speake2003,Kim2010,Behunin2012,Behunin2014,Garrett2015}.
We combine the different sources of uncertainty in order to provide a total estimate of the uncertainty in the comparison between calculations and experiment. 
For our measurements, uncertainty in separation is found to dominate the error at distances < 1\nt{2}0 nm, while the hydrodynamic force dominate\nt{s} the error at separations > \nt{120} nm.

\subsection{The atomic force microscope}

All AFMs contain a microcantilever, a system to control the sample position (typically a piezoelectric transducer), a system to excite the cantilever (piezoelectrically, electrostatically, photothermally, {\it etc}), and a method to detect the motion of the cantilever (optical lever, interferometer, piezoelectric current, {\it etc}). 
In this article, we discuss AFMs up to the level of detail necessary to describe the artifacts present in Casimir force measurements and to discuss strategies to mitigate those artifacts. 
Although the sources and amount of force uncertainty vary from system to system, some sources of uncertainty follow characteristic trends. 
For example, almost all sphere-plate Casimir force measurements rely on the electrostatic force for the estimation of the absolute separation or the calibration of force sensitivity, but its accuracy has only been tested a few times \cite{Kim2008,DeMan2009}.  
	
Because the AFM used here (Cypher, Asylum Research) is very similar to the AFMs used in prior Casimir force measurements, an analysis of the uncertainty the microscope imparts to the measurement helps to predict the uncertainty present in other systems. 
For example, the signals output from each lock-in amplifier (LIA) contain a small offset voltage ($\sim -180$ $\mu$V) that varies over time. 
To track the signal's variation, a null signal is collected at each point and is averaged over time to reduce noise. 
In addition, the driving signal from each direct digital synthesizer (DDS) couples directly into the output of the corresponding LIA. 
	
In the AFM, the piezoelectric transducers actuate the sample and cantilever, as is common to many force measurement procedures \cite{Chang2012,DeMan2009a}. 
The motion of the cantilever is detected by an optical lever, a beam of light reflected off of the cantilever and onto a quad-photodiode \cite{Mohideen1998,VanZwol2008c,Sedighi2016}. 
LIAs then monitor the motion of the cantilever at a several frequencies.
A DDS controls the voltage difference between the probe and sample.
\ntt{A temperature of 303.15 $\pm 0.05$ K is maintained inside the AFM}.

\subsection{Overview of the force measurement method}

\begin{figure}[ht]
	\centering
	\includegraphics[width=.45\textwidth]{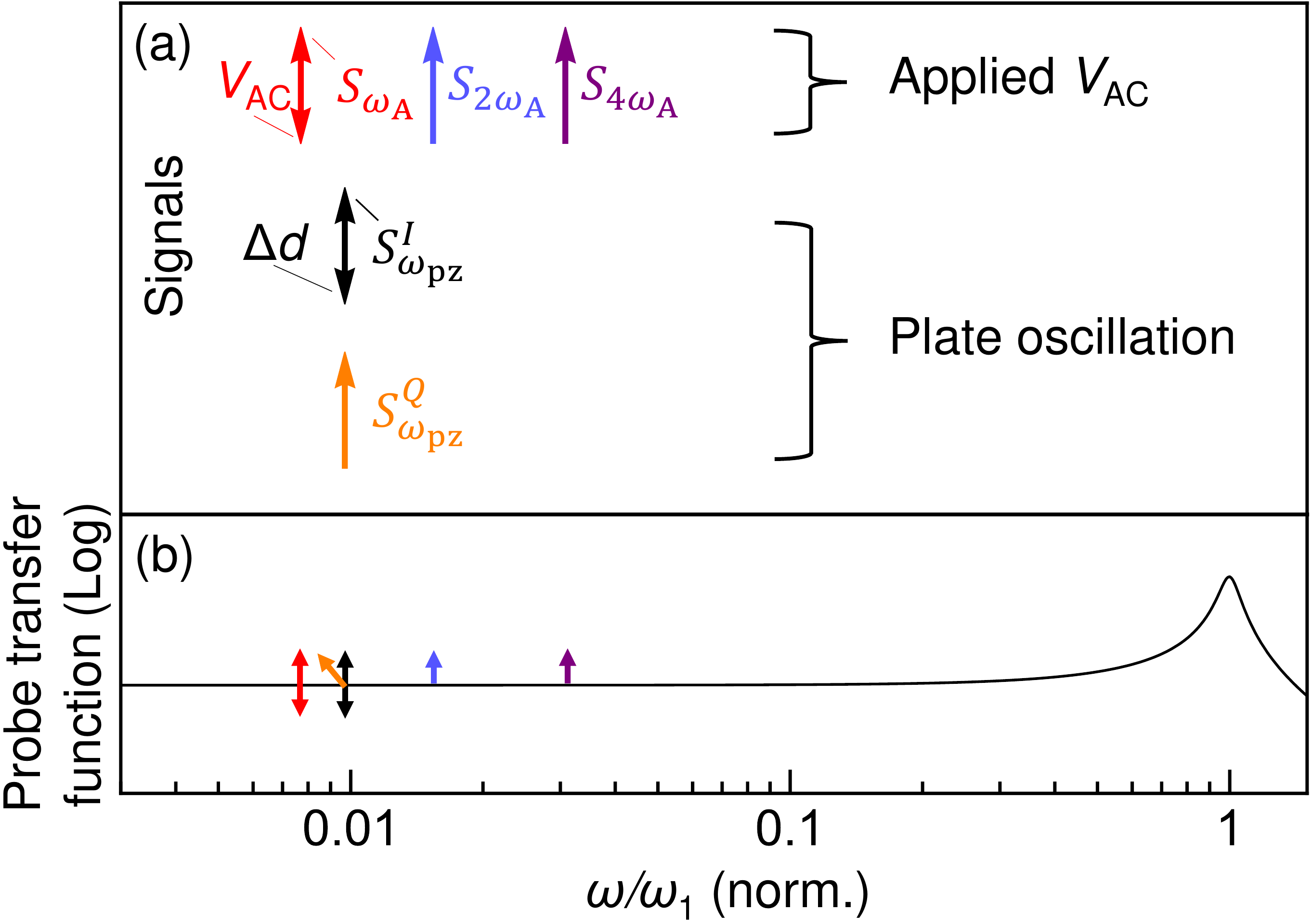}
	\caption[The signals used for the force measurement]{ 
		\nt{(a)} An AC voltage ($V_{AC}$) is applied to the sphere to generate three signals ($S_{\omega_{\text{A}}}$,$S_{2\omega_{\text{A}}}$,$S_{\nt{4}\omega_{\text{A}}}$) that are used to minimize the potential difference, estimate the sphere-plate separation, and calibrate the sensitivity.
        Oscillating the plate drives oscillations of the cantilever that are proportional to the Casimir force (in-phase, $S^{I}_{\omega_{\text{pz}}}$) or hydrodynamic force (quadrature, $S^{Q}_{\omega_{\text{pz}}}$). 
        \nt{(b)} The transfer function describes how forces at different frequencies excite cantilever oscillation. 
        All the frequencies are plotted normalized by the first resonance frequency of the cantilever, $\omega_{1}$.
        Downward arrows represent perturbations applied to the cantilever, upward arrows indicate signals generated by the response of the cantilever, \nt{and the tilted-upward arrow represents an out-of-phase response}.
	}
	\label{fig:cantilevertransfer}
\end{figure}

The force measurement method that we use here follows the phase-separated force modulation method developed by de Man {\it et al.}\ \cite{DeMan2009a,DeMan2010}.
Figure \ref{fig:cantilevertransfer} shows the general scheme for applying and detecting signals.
To detect the force, the plate's position is oscillated.
The in-phase and 90 degrees delayed (quadrature) response of the cantilever are tracked and related to the Casimir and hydrodynamic forces, respectively. 

The electrostatic force between the two surfaces is used both to determine their absolute separation, so that the force versus separation profile can be obtained, and to calibrate the detected Casimir force signal.
The plate is slowly brought towards the sphere \nt{in} discrete \nt{steps}. 
The sphere approaches and retracts from the plate about 30  times in about 13 hours. 
The details of this technique and two other methods for measuring the Casimir force are described in the following sections. 

The interacting surfaces are coated with gold because it is a chemically inert conductor. 
The plate is a silicon substrate coated with 100 nm of gold (e-beam) using 5 nm of Cr for adhesion. 
The sphere is made of hollow glass (Trelleborg SI-100) coated with TiO$_{2}$(10 nm)/Ti(3 nm)/Au(100 nm). 
Both the sample and probe are cleaned with acetone/isopropyl alcohol/DI water and dried with an anti-static air flow prior to the gold deposition.
Two different types of cantilever are used in the measurement: HQ:CSC37/Cr-Au (Mikromasch) cantilevers are used for the analysis leading to Fig. \nt{10 and 11}, while a MLCT-OW-B (Bruker) cantilever is used elsewhere.



\section{Force modulation measurement technique}\label{s:forcemeasurement}

The force modulation (FoM) technique of de Man {\it et al.}\ drives sinusoidal cantilever oscillation with the Casimir force directly by shaking the plate vertically at frequency $\omega_{\text{pz}}$ \cite{DeMan2010}. 
Because the position of the plate varies sinusoidally, so does its velocity, $ v = \partial d / \partial t$.

\begin{figure}[ht]
	\centering
	\includegraphics[width=.45\textwidth]{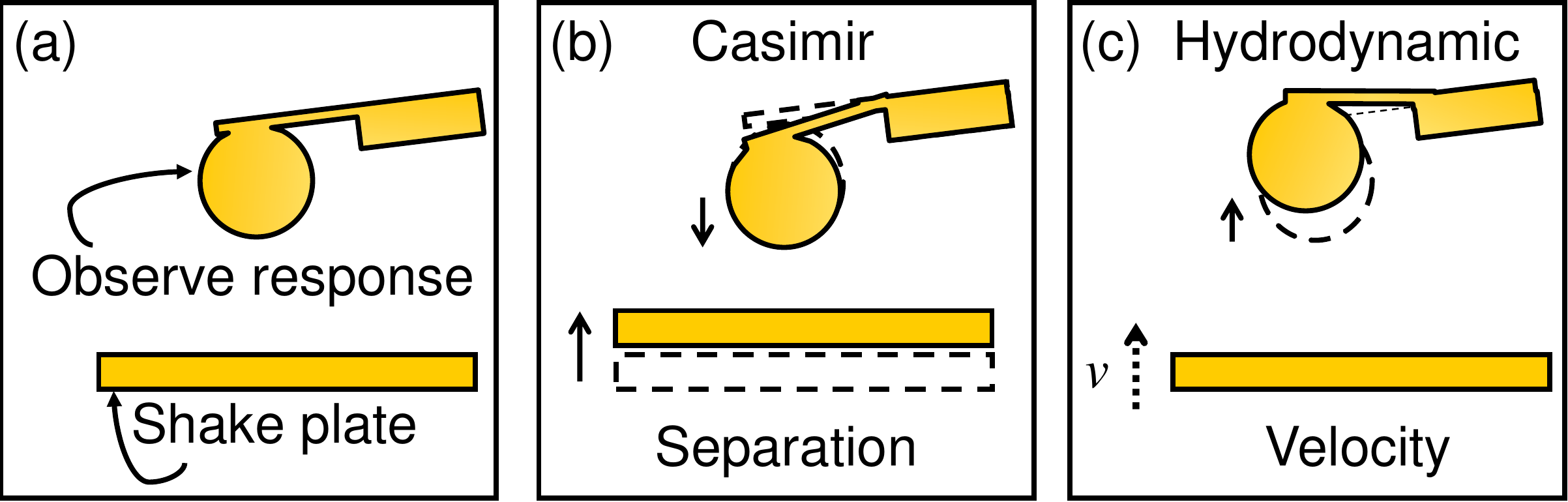}
	\caption[Force modulation method of force detection]{ 
		(a) The sphere-plate force is probed by shaking the plate and observing the response of the probe. 
		(b) Because conservative forces (Casimir, electrostatic) depend on position but not velocity, they bend the probe proportionally to the plate's displacement (in-phase). 
		(c) On the other hand, because the hydrodynamic force is proportional to velocity, the cantilever bending it causes is 90 degrees out-of-phase with the plate's displacement. 
	}
	\label{fig:MeasurementTechnique}
\end{figure}

The response of the cantilever to the moving plate has both an in-phase and a quadrature component (Fig. \ref{fig:MeasurementTechnique}):
\begin{subequations}
	\begin{align}	
		\label{eq:SI}
		S^{I}_{\omega_{\text{pz}}} &= \frac{\gamma}{k}\bigg(\frac{\partial F_{\text{es}}}{\partial d}+\frac{\partial F_{\text{C}}}{\partial d}\bigg)\Delta d,\\
		S^{Q}_{\omega_{\text{pz}}} &= \frac{\gamma}{k}F_{\text{H}}(v),
	\end{align}
\end{subequations}
where $\gamma$ is the optical lever sensitivity (V/m), $\Delta d$ is the shake amplitude of the plate, $k$ is the spring constant, $F_{\text{C}}$ is the Casimir force, $F_{\text{es}}$ is the electrostatic force, and $F_{\text{H}}$ is the hydrodynamic force. 
Derjaguin's proximity force approximation (PFA) is used to compare the measured signal between a sphere and a plate to the Casimir force between parallel plates per unit area, $F_{\text{pp}}$:
\begin{align}
	\frac{1}{R}\frac{\partial F_{\text{C}}}{\partial d} \approx  2 \pi F_{\text{pp}}.
\end{align} 

\subsubsection{Generating the force signal}

During the measurement, the sphere begins about 5 $\mu$m from the plate and approaches it at discrete separations until it reaches a preset minimum.
Then the direction of motion is reversed so that it is similarly retracted from the surface.

At each separation, the measurement is performed in three steps. 
During the first step, an AC voltage, $V_{\text{AC}}$, is applied to the sphere at frequency $\omega_{\text{A}}$ in order to drive the cantilever at frequencies $\omega_{\text{A}}$ and $2\omega_{\text{A}}$, while the plate is grounded. 
The cantilever's response to the applied voltage is detected with LIA A.
A feedback loop uses the signal at $2\omega_{\text{A}}$ to control $V_{\text{AC}}$ in order to maintain a constant amplitude set point, $A_{\text{set}}$. 
A second feedback loop applies a slowly varying voltage, $V_{\text{DC}}$, to the sphere in order to minimize the signal at $\omega_{\text{A}}$, which in turn minimizes the potential difference between the sphere and the plate. 
The electrostatic force generated by $V_{\text{AC}}$ has a large signal-to-noise ratio, so it is used to \nt{account for the change in $d_{0}$, also called drift, over the course of the measurement.}

The force measurement is performed in the second step at each separation. 
First, the oscillating voltage $V_{\text{AC}}$ is turned off.
Second, $V_{\text{DC}}$ is set to -$V_{0}$, its average value over the first step, to mitigate the electrostatic force.
Third, while a piezo oscillates the plate, the response of the cantilever is detected by the optical lever and recorded by LIA B. 

During the third step, the piezo continues to oscillate the plate while $V_{\text{DC}}$ is discretely varied across its force-minimizing value in order to determine the electrostatic force gradient. 
The $V^{2}$ dependence of the electrostatic force causes the signal versus voltage curve to take the shape of a parabola.
The range of the voltage sweep is chosen so that the total signal variation of the parabola (and thus the sensitivity) remains approximately constant at every separation. 
The second $d$-derivative of the capacitance ($C''$) is calculated from the curvature of each voltage parabola.
In turn, $C''$, discussed in more detail below, is used to determine the tip-sample separation. 
\nt{The electrostatic force gradient allows us to determine the separation more accurately because the gradient changes more quickly with separation than the force itself, is measured through the same channel as the Casimir force gradient, and is less susceptible to second-order oscillation}.
Below 110 nm, the third step is stopped to prevent the electrostatic force from causing the tip to jump to contact.
The measurements here use $\omega_{\text{pz}}/2\pi$=211 Hz and $\omega_{\text{A}}/2\pi$=77 Hz.

\subsubsection{Ratcheting}\label{sec:ratchet}

Because the Casimir force signal is proportional to $\Delta d$ (Eq. \ref{eq:SI}), increasing the shake amplitude improves the signal-to-noise ratio and enables observations of the Casimir force at larger separations. 
However, using a larger amplitude both limits the minimum achievable separation and can lead to a systematic, but well understood, overestimate of the Casimir force $\propto\Delta d^{3}$ \cite{Lamoreaux2015}.

\begin{figure}[ht]
	\centering
	\includegraphics[width=.44\textwidth]{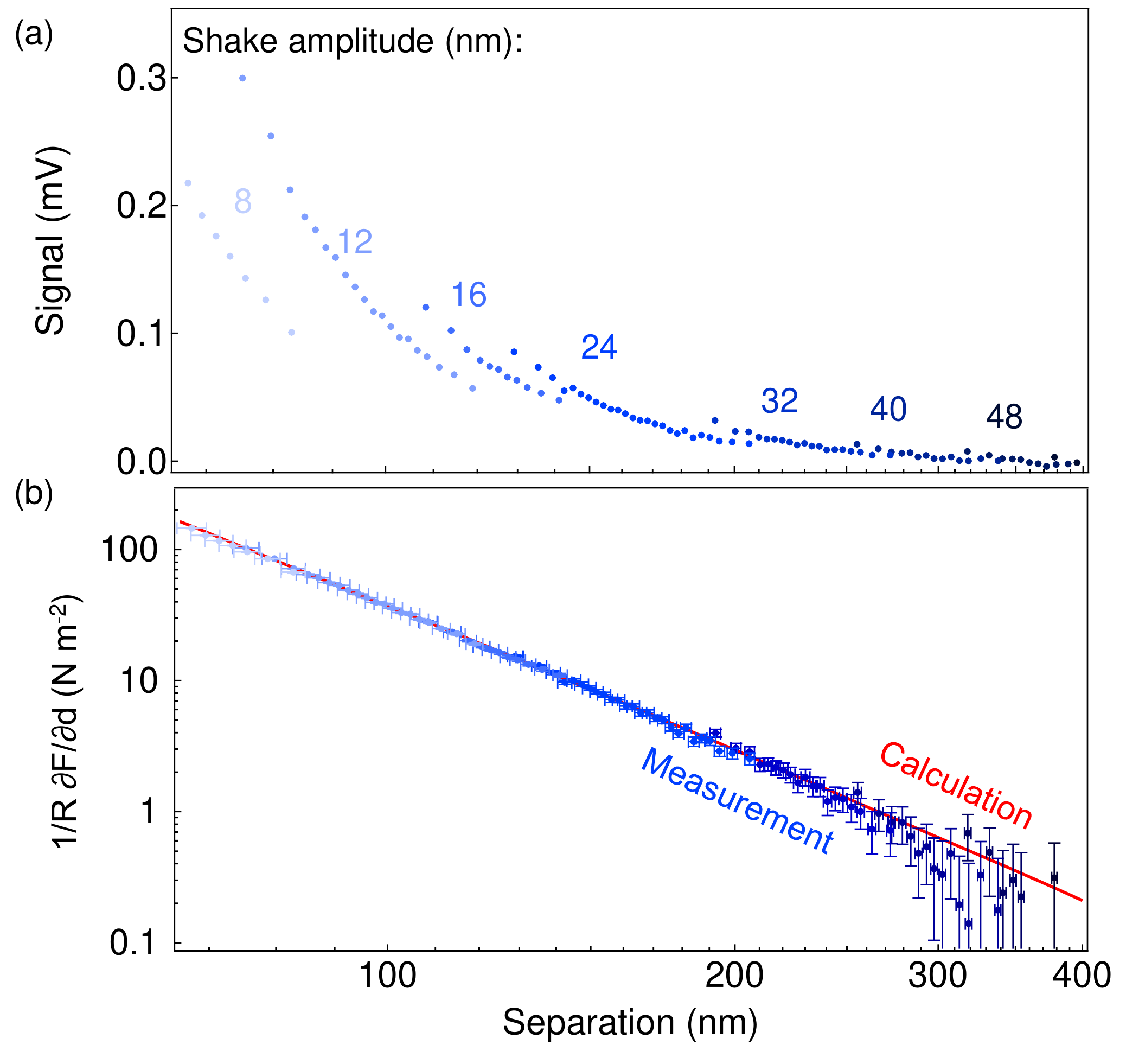}
	\caption[Ratcheting in a force measurement]{ 
		(a) The shake amplitude varies with separation to increase the sensitivity of the force modulation measurement technique, while also avoiding errors associated with the strong nonlinearity of the force. 
		(b) The data collected at each shake amplitude are combined for the final estimate of the force gradient. 
		The data are shown binned into groups of $\approx$50 individual measurements. The red line shows the calculated Casimir force gradient
	}
	\label{fig:Ratcheting}
\end{figure}

To maximize the signal while mitigating the errors associated with large shake amplitudes, a ratcheting technique is introduced.
Far from the surface, the plate oscillates with $\Delta d$ $\approx$ 48 nm. 
To minimize the systematic error from the shake amplitude, $\Delta d/d \equiv \chi$ < 0.15 is maintained throughout the experiment. 
The systematic error from the $\Delta d$ as a fraction of the force gradient is then 2.5$\chi^{2}$, calculated from equation 6 of \cite{Lamoreaux2015} by assuming the typical $d^{-4}$ dependence of the sphere-plate Casimir force gradient.
Thus, here the error is kept below 6\% of the force gradient.
For example, once the plate reaches 320 nm from the surface, the shake amplitude decreases to 40 nm. 
Likewise, when \nt{the} sphere reaches 267 nm separation, the amplitude drops to 32 nm. 
The process repeats so long as the tip is approaching the surface. 
When the plate is retracting, the process is reversed. 
We call the technique `ratcheting' because, on approach, the shake amplitude only decreases and, during retraction, it only increases (Fig. \ref{fig:Ratcheting}). 

\section{Fundamental limits to the measurement range}\label{sec:thermal}
Understanding the fundamental limits to \nt{FoM} Casimir force measurements helps to frame the effects of other sources of uncertainty.
\nt{In this section the perfect conductor approximation for the Casimir force is used, because it permits analytic results. 
For real materials, the force approaches a $d^{-3}$ rather than a $d^{-4}$ power law at short separations.
The separation at which this transition begins depends on the particular material.}

Jump-to-contact (JTC) limits how close to the surface Casimir probes can approach, and for measurements in which the shake amplitude is much less than the separation ({\it e.g.} for deflection measurements, or the force modulation measurements discussed here), the criterion for JTC is $k < \partial F/ \partial d$ \cite{Burnham1989,Giessibl2003}. 
The minimum possible separation is then limited by the JTC, so that when the dominant force is the Casimir force \cite{Casimir1948a},
\begin{align}
	\label{eq:JTC}
	d_{\text{min}} \nt{\approx} \bigg(\frac{\hbar c \pi^{3}}{120}\frac{R}{k}\bigg)^{1/4}.
\end{align}
A typical probe (table \ref{table:typical_cantilever}) should be able to measure as close as 43 nm from the surface, which approximately agrees with experiment. 
Because of the $d^{-3}$ power law in the force, Eq. \ref{eq:JTC} is fairly insensitive to $k$. 
For example, if $k$ is increased by a factor of ten to 1 N/m, $d_{\text{min}}$ only decreases to 24 nm, less than a factor of two. 

Thermal noise limits the furthest separation at which the force can be measured. The minimum detectable force for the FoM method described below is $F_{\text{min}}= k n_{\text{d}} \sqrt{B}/\gamma$, where $n_{\text{d}}$ (V Hz$^{-1/2}$) is the noise amplitude density at the detector and $B$ is the detection bandwidth. 
In the experiments discussed here, the $n_{\text{d}}$ is dominated by the detector, but the fundamental limit to sensitivity is the cantilever's thermal motion. 
Because the oscillation frequency is much less than the resonant frequency, only the first eigenmode of the cantilever is considered.  When thermal noise is dominated by the cantilever's motion \cite{Heer1972}
\begin{align}
	n_{\text{d}} = 2\gamma\sqrt{\frac{k_{\text{B}}T}{k\omega_{1}Q}}.
\end{align}
Then the minimum detectable force is
\begin{align}
	\label{eq:Minimum_detectable_force}
	F_{\text{min}}= 2 \sqrt{\frac{k k_{\text{B}} T}{\omega_{1} Q }} \sqrt{B},
\end{align}
which, with the properties of a typical cantilever (table \ref{table:typical_cantilever}) and $B$ = 1 Hz, is about 20 fN.
However, our technique measures the force spatial derivative rather than the force.
The minimum detectable force gradient, when the cantilever is driven by $\partial F/\partial d \Delta d$ is
\begin{align}
	\label{eq:Minimum_detectable_forceGRAD}
	F'_{\text{min}} = 2 \sqrt{\frac{k k_{\text{B}} T}{\omega_{1} Q }} \frac{\sqrt{B}}{\Delta d},
\end{align}
where $\Delta d$ is the oscillation amplitude of the plate.
The maximum separation is found by finding the separation at which the predicted force gradient equals the minimum detectable force gradient 
\begin{align}
	\label{eq:max_separation}
	d_{\text{max}} = \bigg(\frac{\hbar c\pi^{3}R}{240}\frac{\Delta d}{\sqrt{B}}\sqrt{\frac{\omega_{1}Q}{kk_{\text{B}}T}}\bigg)^{1/4}.
\end{align}
The appearance of $\Delta d$ suggests that it is possible to increase $d_{\text{max}}$ arbitrarily, but $\Delta d$ must always be significantly less than $d$ so that the sphere does not hit the surface and to avoid systematic errors associated with the non-linearity of the force \cite{Lamoreaux2015}. 
The shake amplitude must be bounded to mitigate systematic artifacts, $\Delta d$ < $\chi d$ (section \ref{sec:ratchet}), so that
\begin{align}
	d_{\text{max}} < \bigg(\frac{\hbar c\pi^{3}R}{240}\frac{\chi}{\sqrt{B}}\sqrt{\frac{\omega_{1}Q}{kk_{\text{B}}T}}\bigg)^{1/3}.
\end{align} 
For typical cantilevers with $B = 1$ Hz and $\chi = 0.15$, the force detection is limited to separations $d_{\text{max}} < 1.4$ $\mu$m (table \ref{table:typical_cantilever}). 
Note that the effective power law of the sensitivity falls from $d^{-4}$ to $d^{-3}$ when $\Delta d$ is allowed to vary, showing how ratcheting increases the range over which the force gradient may be measured, relative to the case when $\Delta d$ is constant.

Measurements of the Casimir force by the FoM technique of de Man {\it et al.}\ are thus fundamentally limited to separations between $\sim$ 40 nm to 1.4 $\mu$m, or about one and a half orders of magnitude, which is comparable to the largest ranges probed by previous measurements \cite{Lamoreaux1997,Sushkov2011a}. 
Using several probes with varying $R$ and $k$ may increase the range.
The next section discusses the sources of uncertainty that prevent measurements from achieving the range set by fundamental limitations.

\begin{table*}[t]
	\caption{Typical probe properties}
	\centering
	\begin{tabular}{ c  c  c  c  c c c}
	 $\frac{\omega_{1}}{2\pi}$  (kHz)& L ($\mu$m) & W ($\mu$m) &R ($\mu$m) & $k$ (N/m) & $1/\gamma$ (nm/V) & Q \\
		\hline
		 10 & 250 & 33 & 40 & 0.1 & 700 & 100\\
	\end{tabular}
	\label{table:typical_cantilever}
\end{table*}

\section{Calibration and separation determination}\label{calibration}

The calibration and separation determination in Casimir force measurements are most often performed with the electrostatic force, although the hydrodynamic force has been used as well in liquids, where Debye screening affects the electrostatic force \cite{Munday2008b,Cunuder2018}.
In the low Reynolds number limit, the hydrodynamic force is proportional to $d^{-1}$, so it might also be possible to use it to estimate the tip-sample separation in air, as it has been used in liquid \cite{Munday2007}. 
The difficulties with using the hydrodynamic force are twofold: (1) the hydrodynamic force is nearly two orders of magnitude weaker in air than in water, so the signal-to-noise ratio of its detection is smaller\nt{, although it could be increased by increasing $\omega_{\text{pz}}$}, and (2) the slip length at ambient pressures is quite large (\nt{estimates range from} 60 nm \cite{Laurent2011a} \nt{to 118 nm \cite{Maali2008}}) and, while it can be included in the fit, the extra free parameter further reduces the accuracy of the separation estimation.
Because $\sim$60 nm of separation uncertainty would prevent theory-experiment comparison, calibration with the electrostatic force is the focus of this section. 

\subsection{\nt{From t}he electrostatic force}
The electrostatic force between a plate and a sphere is
\begin{align}
	F = \frac{C'(V+V_{0})^{2}}{2},
	\label{eq:Sphere_place_capacitance}
\end{align}
where $V$ is the applied potential between the plate and the sphere, $V_{0}$ is the minimizing potential, and $C'=\partial C/\partial d$, where $C$ is the sphere-plate capacitance,
\begin{align}
	\label{eq:full_cap}
	C' &= 2\pi \epsilon_{0} R \sum_{n=1}^{\infty}\frac{\coth(\alpha)-n\coth(n\alpha)}{\sinh(n\alpha)},
\end{align}
and $\alpha$ is defined by the equation $\cosh(\alpha)=1+d/R$ \cite{Decca2003}.
Note that because $C'<0$, the electrostatic force is attractive.
The voltage applied to the probe has two components: $V_{\text{AC}}$ and $V_{\text{DC}}$, so that the total voltage between the surfaces is $V=V_{\text{AC}}\cos(\omega_{\text{A}}t)+ V_{\text{DC}}+V_{0}$. 
We can separate the electrostatic force into three terms
\begin{align}
	F_{\text{es}} = F_{\text{DC}} + F_{a} + F_{b},
	\label{eq:electrostaticforce}
\end{align}
where the individual forces are separated according to the frequency of the applied voltage
\begin{subequations}
	\begin{align}
		\label{eq:electrostatic_0freq}
		F_{\text{DC}} &= \frac{C'(d)}{2}\big( (V_{\text{DC}}+V_{0})^{2}+\frac{V_{\text{AC}}^{2}}{2}\big),\\
		\label{eq:omegaV}
		F_{a} &= C'(d)V_{\text{AC}}(V_{\text{DC}}+V_{0})\cos(\omega_{\text{A}}t), \\
		\label{eq:omega2V}
		F_{b} &= \frac{C'(d)}{4}V_{\text{AC}}^{2}\cos(2\omega_{\text{A}}t),
	\end{align}
\end{subequations}
where it is noted that $C'$ itself depends on $d$, which varies with time because of both the oscillations of the plate and the cantilever, which is why the forces are labeled with $a$ and $b$ rather than frequencies. 

Signals generated at the frequencies $\omega_{\text{A}}$ and $2\omega_{\text{A}}$ are critical for tracking relative changes to the potential difference and separation.
The signal at $\omega_{\text{A}}$ is generated by $F_{a}$ and is used as the input to the feedback loop that measures $V_{0}=-V_{\text{DC}}$, akin to the loop used in Kelvin probe force microscopy \cite{Nonnenmacher1991}. 
The force $F_{b}$ generates the signal at $2\omega_{\text{A}}$: $S_{2\omega_{\text{A}}}\approx\gamma\frac{C'}{4}V^{2}_{\text{AC}}/k$. 
Although there are systematic artifacts in determining $V_{0}$ with the signal at $\omega_{\text{A}}$, the low noise level permits the tracking of the contact potential difference over time \cite{Melin2011,Barbet2014,Polak2014}. 
Likewise, the measurements of $C'$ from $2\omega_{\text{A}}$ have a high signal-to-noise ratio, which makes them useful for correcting for drift in position and sensitivity between different approach/retract runs (see section \ref{sec:separation}). 

To estimate the relative separation and sensitivity, the measured values of $S_{2\omega_{\text{A}}}$ are fit to a function of the form
 \begin{align}
 	\label{eq:Cfit}
	\frac{S_{2\omega_{\text{A}}}}{V^{2}_{AC}} = \frac{\kappa}{2R}C'(d_{\text{pz}}-d_{0},R),
 \end{align}
in which $d_{\text{pz}}$ is the position of the plate relative to a reference height, and the two free parameters are the sensitivity ($\kappa = \gamma R/2k$) and the absolute position offset ($d_{0}$).
Fitting separates the two parameters $d_{0}$ and $\kappa$. 
The relative piezo displacement is typically measured accurately, for example, by a linear differential transformer, so that the electrostatic force can be fit assuming that relative displacements over a measurement are exact, and only $d_{0}$ is unknown.
Both $\gamma$ and $k$ are assumed to be frequency-independent because the frequencies used are much lower than the resonant frequency of the cantilever. 
\ntt{The high resonant frequency of the cantilever is enabled through the use of a hollow, rather than solid, glass sphere.}
Any drift in $V_{0}$ or sensitivity across time is calibrated for using the $C'$ measurements.

\subsection{\nt{From the} electrostatic force gradient}

\begin{figure}[ht]
	\centering
	\includegraphics[width=.4\textwidth]{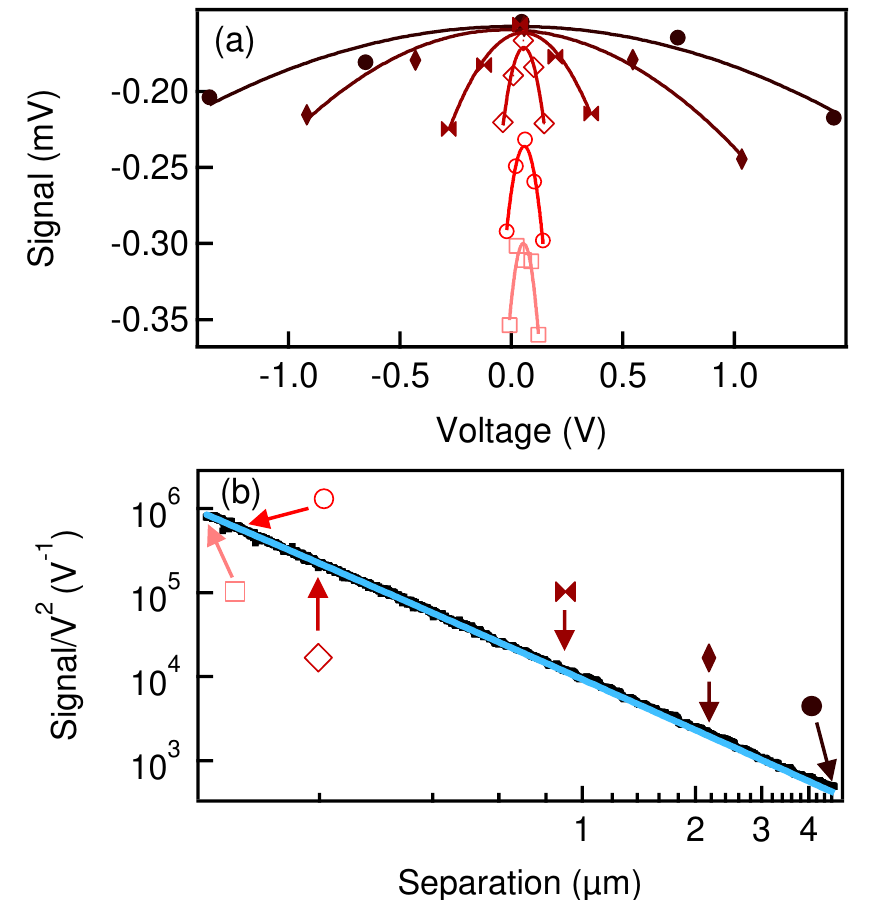}
	\caption[parabolas]{ 
		(a) The total force gradient signal as a function of applied voltage at several different separations \nt{(eq. \ref{eq:Cppsig})}, each noted by a symbol in (b), where the measured force gradient (black dots) is shown with the fit used to determine the absolute separation (blue line). 
	}
	\label{fig:ForceGradient}
\end{figure}

We determine the absolute position and sensitivity from measurements of the gradient of $F_{\text{DC}}$, measured through the same channel as the Casimir force (Fig. \ref{fig:ForceGradient}).
The plate is oscillated, as it is when the Casimir force signal is generated, so that the force gradient between the sphere and the plate is measured while $V_{\text{AC}}=0$. 
Simultaneously, $V_{\text{DC}}$ is slowly varied across the -$V_{0}$ measured in the first electrostatic step. 
Because there is no AC voltage, there is no AC coupling to the piezo.
The $V^{2}$ dependence of the electrostatic force causes the resulting signal to be a parabola, whose curvature can be related to $C''$:
\begin{equation}
	S_{\text{DC, pz}} = \frac{\epsilon_{0}\pi}{2}C''(V_{\text{DC}}+V_{0}^{C''})^{2} \frac{R \gamma \Delta d}{k},
	\label{eq:Cppsig}
\end{equation}
where the superscript on $V_{0}^{C''}$ shows that it is the force gradient minimizing voltage. 
Because the signal-to-noise ratio of the force gradient measurement is much lower than the force measurement, it is not used \nt{to} estimate the minimizing voltage or separation during the measurement \nt{for individual approaches.} 
\nt{However, in the final analysis $C''$ is used to calculate separation and the minimizing voltage, because it is much less susceptible to artifacts than $C'$. 
If the measurement of $C'$ did not contain the systematic effects discussed below, $C''$ could be calculated directly from it. 
However, in practice, calculating $C''$ from $C'$ directly, rather than determining it from a second measurement, amplifies the effect of the artifacts discussed below.}

To compensate for the low signal-to-noise \nt{ratio} of the force gradient measurement, the relative positions of many runs are aligned before fitting the parabola\nt{, as discussed below in the section \ref{drift}.}
Once the curvature of the parabolas as a function of position is determined, it is fit to $C''$ to determine the absolute position and sensitivity. 
By measuring the electrostatic force through the same channel as the Casimir force, any non-idealities in the piezo actuation of the plate appear in the electrostatic calibration as well as the Casimir force, which helps with diagnosing experimental problems, such as uncertainty about the shake amplitude, $\Delta d$ of the plate \cite{Eerkens2017}.

\subsection{Determining spring constant and optical lever sensitivity}\label{sec:calibration}

In our experiment, higher harmonics driven by the non-linearity of the electrostatic force are used to separate the spring constant $k$ from the optical lever sensitivity, $\gamma$ (which are combined to give $\kappa$ in Eq. \ref{eq:Cfit}). 
To calculate them, we expand $C'$ to 1st order in a Taylor series around the time-averaged separation
\begin{align}
	\label{eq:CpExpansion}
	C'(t) &= C'(d)+C''(d) A \cos(2\omega_{\text{A}}t) + ... 
\end{align}
Then, inputting Eq. \ref{eq:CpExpansion} into Eq. \ref{eq:omega2V}, forces at higher frequencies are found
\begin{align}
	F_{b} = &\frac{C'}{4} V_{\text{AC}}^{2}\cos(2\omega_{\text{A}}t)\\
	&+\frac{ C'C''}{16k}V_{\text{AC}}^{4}\cos^{2}(2\omega_{\text{A}}t).\notag
\end{align}
The second term can then be expanded, so that the electrostatic force on the cantilever at frequency $4\omega_{\text{A}}$, up to first order, is
\begin{align}
	F_{4\omega_{\text{A}},1} = \frac{ C''C'}{32k}V_{\text{AC}}^{4}\cos(4\omega_{\text{A}}t).
	\label{eq:4f_signal}
\end{align}
Note, when the PFA for the capacitance is inserted into Eq. \ref{eq:4f_signal}, the strength of the force is consistent with the calculation of de Man {\it et al.}\ in which the PFA is assumed from the beginning \cite{DeMan2010}:
\begin{align}
	F_{4\omega_{\text{A}},1} = - \frac{\pi^{2}\epsilon_{0}^{2}R^{2}}{8kd^{3}}V_{\text{AC}}^{4}\cos(4\omega_{\text{A}}t).
	\label{eq:4f_signalPFA}
\end{align}
Because $F_{4\omega_{\text{A}},1}$ depends on $k$, independent of $\gamma$, the signal that it drives, $S_{4\omega_{\text{A}}}$, can be used to separate the two parameters. 
To do so, the electrostatic force is driven with $V_{\text{AC}}$ = 8 V on approach, so that both the $S_{2\omega_{\text{A}}}$ and $S_{4\omega_{\text{A}}}$ signals are generated.
 
\nt{Within one approach,} determination of $d_{0}$ is performed with the $S_{2\omega_{\text{A}}}$ signal.
Using the $d_{0}$ found from the first fit, the $S_{4\omega_{\text{A}}}$ signal is fit to Eq. \ref{eq:4f_signalPFA}. 
Then $S_{4\omega}$ is used to separate $k$ and $\gamma$ as
\begin{subequations}
		\begin{align}
		k &= \frac{\epsilon_{0}\pi R}{4}\frac{\kappa}{S_{4\omega}}, \\
		\gamma &= \frac{\kappa^{2}}{2S_{4\omega}},
	\end{align}
\end{subequations}	
where $\kappa$ comes from the $C'$ fit (Eq. \ref{eq:Cfit}).
It is not necessary to split $\kappa$ into $k$ and $\gamma$ to perform measurements, but doing so permits the comparison of our calibration to other calibration methods.
In our previous measurements \cite{Garrett2018}, we ran a large-$V_{\text{AC}}$ calibration every other run, but now, we only do so at the end of a set of runs. 

\section{Determining measurement uncertainties}

\subsection{Characteristics of different uncertainties}

\nt{To determine the total uncertainty in the measurement, we divide it} into several categories. 
\nt{All uncertainties are reported in percent of the measured force. 
However, separation uncertainty is first reported as a range of separations before being converted into a percent uncertainty.}

First is the calibration uncertainty, which includes uncertainty in the measurements of $k$, $R$, and $\gamma$ and  uncertainty in the calibration of the piezo actuation $\Delta d$ (section \ref{sec:calibration}). 
Because a number of different methods of calibrating AFM cantilevers have been developed, comparing the results of different methods is one way to characterize their uncertainty \cite{Burnham2003}.
Techniques used to estimate $k$ of an AFM cantilever include fitting the cantilever's thermal motion to a Lorentzian,  measuring the change in the cantilever's resonance frequency when it is used to pick up particles of a known mass, or measuring the response to a known radiation pressure \cite{Hutter1993,Cleveland1993,Weld2006a}.
The effect of the added mass on `colloidal probes' (like the probes used here) changes the correction factor used for thermal calibration and affects the cantilever mode shapes \cite{Laurent2013a}.
While variation between different techniques for calibrating $k$ can be as large as 17\% \cite{Burnham2003}, similar electrostatic calibration experiments on colloidal probes suggest that the error in $k$ from using electrostatic calibration is about $5\%$ \cite{Chung2009}. 
\nt{Thus,} a calibration uncertainty \nt{of} $5\%$ \nt{is used}. 

Uncertainty in the absolute position of the sample relative to the probe is one of the most problematic sources of error in Casimir force measurements because of the strong separation dependence of the force. 
The uncertainty in the measured $F'$ is calculated from the uncertainty in the position ($\pm2$ nm) multiplied by $F''$ (section \ref{sec:separation}). 
Note that although the force is measured with no less precision at separations below 120 nm, uncertainty in the separation makes comparison with theory less viable.
\nt{The calculation of the uncertainty in position is presented below.}

Interference \nt{of the optical lever is} a major source of uncertainty in the measurements. 
The interference varies in both phase and magnitude between the different spheres. 
The magnitude of the interference is estimated by the technique described above in section \ref{sec:interference}. 

\nt{The hydrodynamic force present in the Casimir force channel of the measurement in the limit of a small phase offset is $F_{H} \sin(\Delta\theta_{\text{ref}})$, where $\Delta \theta_{\text{ref}}$ is the difference between the reference phase of the lock-in amplifier and the response of the cantilever to the force modulation. 
Below in } section \ref{sec:hydrodynamic}\nt{, $\Delta \theta_{\text{ref}}$ is determined}. 
\nt{Then t}he uncertainty originating from the hydrodynamic force is determined by multiplying the measured hydrodynamic force by $\sin(\Delta\theta_{\text{ref}})$. 		
The magnitudes of both the hydrodynamic force and interference depend significantly on the particular probe used for the measurement.

\nt{Stochastic} noise is estimated by dividing the standard deviation of the \nt{force gradient} data within a small separation range ($\sim$ 2 nm) by the square root of the number of data points collected within that range. 
Here, the stochastic noise comes primarily from the photodetector, but there is always some stochastic noise due to the thermal motion of the cantilever.

The electrostatic force is present because of an artifact in the minimizing voltage detected by the Kelvin probe feedback loop (section \ref{sec:voltageuncertainty}).
\nt{AC coupling causes $V_{0}$ to appear to vary with distance, even if there are no patch potentials.} 
The AC coupling that would cause the \nt{entire separation-dependence of the} measured $V_{0}$ \nt{is calculated \cite{Barbet2014}}. 
\nt{Because AC coupling causes the measured $V_{0}$ to be offset from the actual $V_{0}$, the residual electrostatic force from the estimated $V_{0}$ discrepancy determines the uncertainty due to the electrostatic force.} 

\subsection{Uncertainty in separation determination}\label{sec:separation}

Because the electrostatic force is used to determine the separation between the sphere and the plate, any effects that lead the measured electrostatic force to deviate from the expected form of the electrostatic force lead to an error in the determination of the absolute separation, $d_{0}$. \
Below, we list several effects that can contribute to error in the separation, discuss ways to control them, and quantify the error that they impart.

\subsubsection{Electrostatic approximations and fitting}

\begin{figure}[ht]
	\centering
	\includegraphics[width=.4\textwidth]{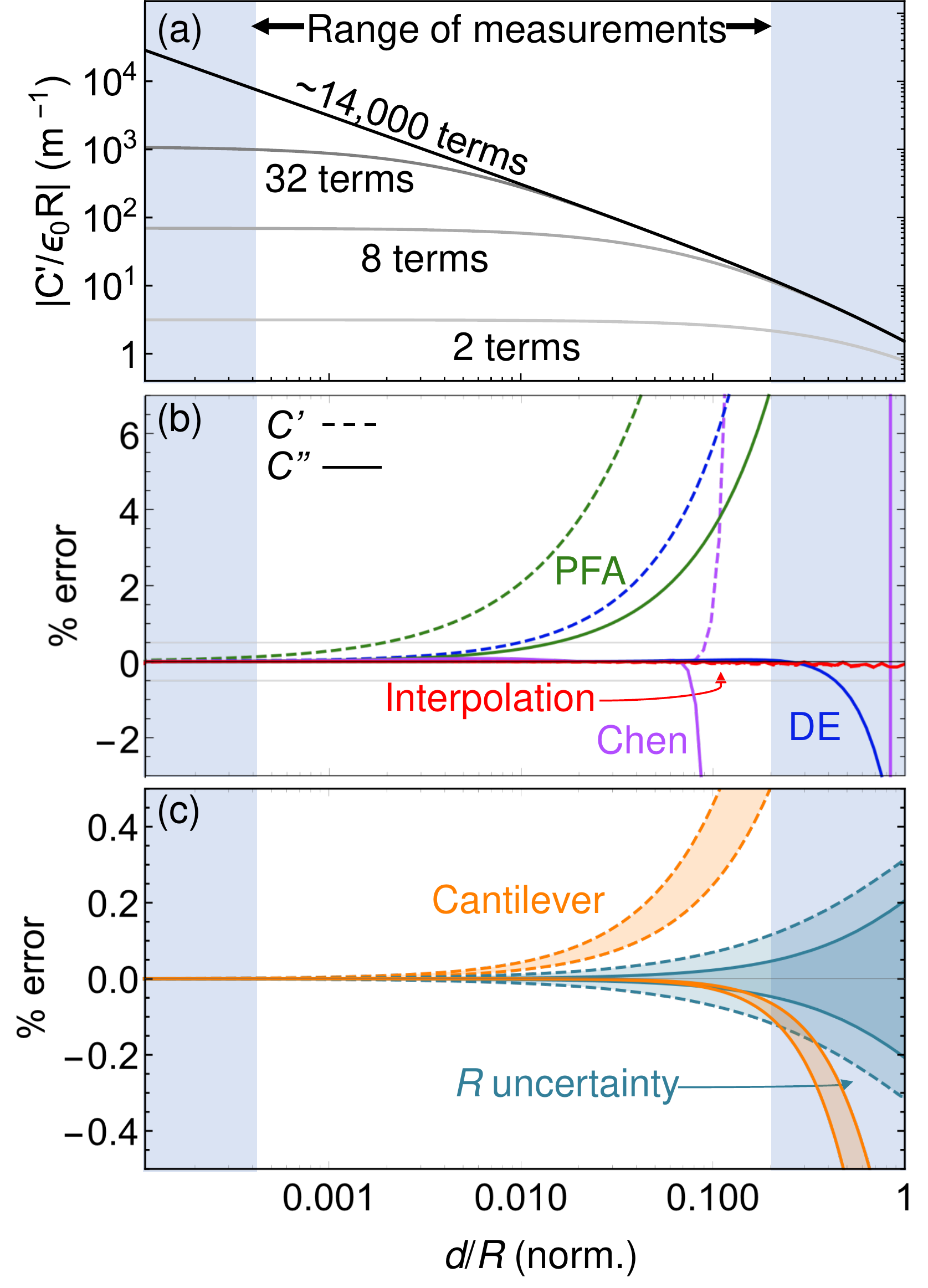}
	\caption[Several approximations for the capacitance gradient]{ 
		(a) The full sphere-plate capacitance gradient ($C'$) infinite sum approximated by several truncated sums. The unshaded central region shows the range of data pertinent to the Casimir force measurements discussed in this article. (b) The proximity force approximation (PFA, green) deviates from the exact solution at large $d$; the approximations of Chen {\it et. al.} \cite{Chen2006} (purple) and the derivative expansion (DE, blue) \cite{Fosco2012} reduce the error somewhat. Logarithmic interpolation (red) shows even less error. For all approximations except Chen's, the $C''$ signal causes less systematic error than the $C'$ signal. (c) Uncertainty in the sphere radius and the contribution of the cantilever itself both affect $C'$ less than using the PFA ($\Delta R \approx$0.7\%, Fig. \ref{fig:Roundness}). 
	}
	\label{fig:SphereCapacitance}
\end{figure}

The computational demands of Eq. \ref{eq:full_cap} have caused several approximations to be used, the most prominent of which is the proximity force approximation (PFA) for $C'$ (Fig. \ref{fig:SphereCapacitance}).
\nt{Moreover, Eq. \ref{eq:full_cap} is exact for an ideal sphere-plate system, but incorporating imperfections such as roughness or a water layer into it is difficult.}
The proximity force approximation (PFA) describes the force between two curved surfaces as the sum of pairwise plate-plate forces. 
To investigate the effects of roughness, long-range sphere deformations and a water layer on the measurement, the effects are calculated using the PFA and then used to estimate error as discussed below. 
Most previous Casimir force measurements have used the PFA or other approximations to Eq. \ref{eq:full_cap} to ease computational difficulties.
Fig. \ref{fig:SphereCapacitance} shows several different implementations of the PFA. 
For a sphere and a plate the approximation gives
\begin{align}
	C_{\text{PFA}}' =-\frac{2\pi\epsilon_{0}R}{d}.
	\label{eq:PFA_cap}
\end{align}

However, the computational difficulty of the infinite sum is not in the evaluation of the sum itself, but rather that the sum must be recalculated for every iteration of the fitting procedure. 
If, instead, the force is calculated once for a number of points, the data for those points can be saved and interpolated for later fit iterations. 
The simplest interpolation, interpolating points linearly, tends to slightly overestimate the force.
Because $C'$ is approximately linear on a log-log plot, interpolating $\log(C')$ vs $\log(d/R)$ gives better accuracy.
It deviates from the exact $C'$ by less than 0.5\% over the whole range of the fitting, which is less than any of the previous approximations, even though the full Eq. \ref{eq:full_cap} sum is only calculated for 43 separations. 
Because the interpolation itself is limited to the values between the minimum and maximum value of $d/R$, it can be helpful for fitting to use Eq. \ref{eq:PFA_cap} for separations below the lowest interpolated value and the $n=1$ term of Eq. \ref{eq:full_cap} for separations above the largest interpolated value. 
Then, the function is well defined \ntt{anywhere that the fitting algorithm may need to evaluate it}. 
Because it is not necessary to use an approximate form of Eq. \ref{eq:full_cap}, approximations do not necessarily impart any separation uncertainty. 
However, we note that it is suspected that Eq. \ref{eq:full_cap} itself is an approximation at some level \cite{Lamoreaux2011b}, and a deviation from it has been reported for separations < 1 nm \cite{Weymouth2012}.

\subsubsection{Sphere radius}

\begin{figure}[ht]
	\centering
	\includegraphics[width=.48\textwidth]{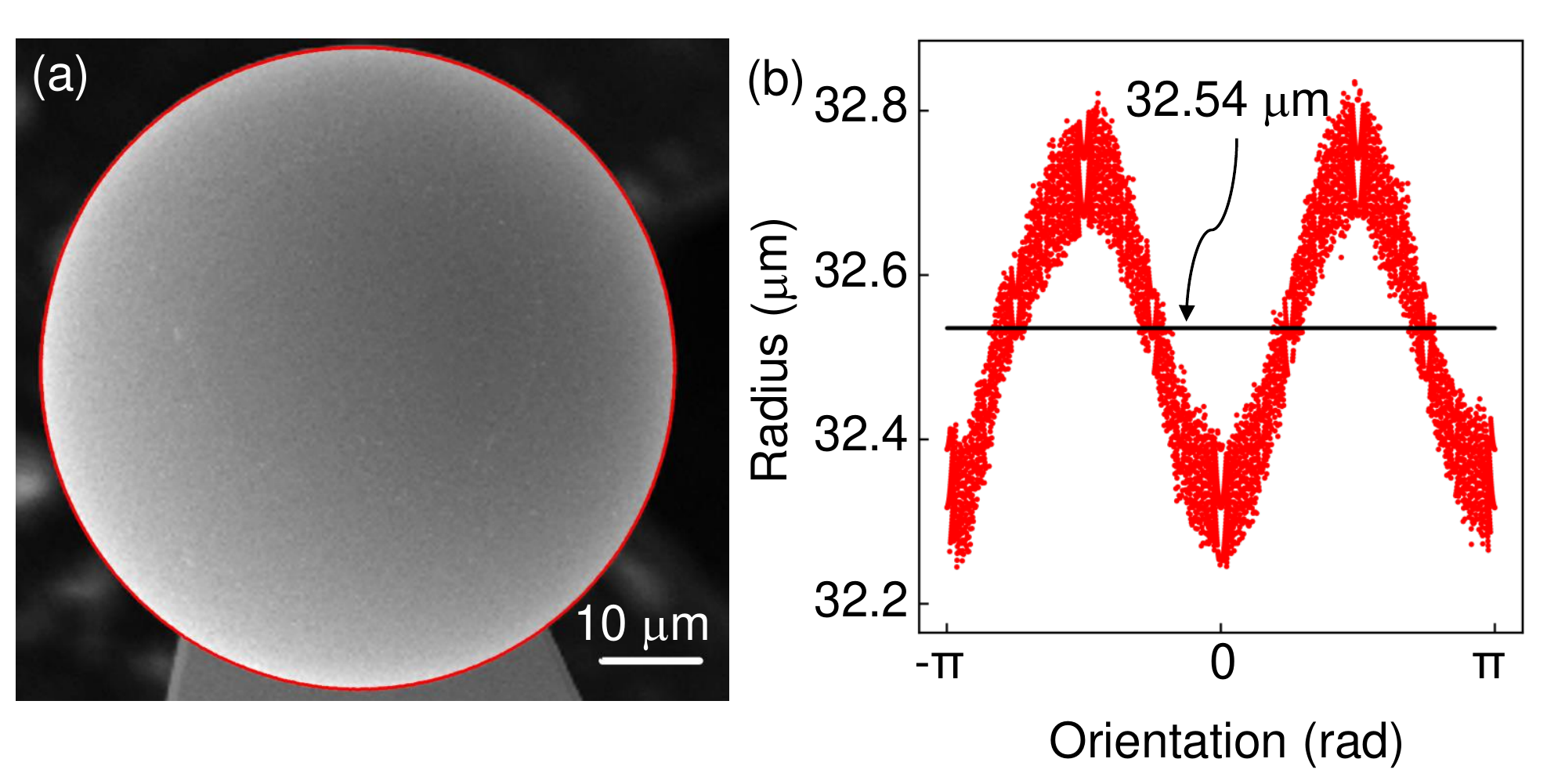}
	\caption[Several approximations for the capacitance gradient]{ 
		(a) The perimeter of a sphere is identified through a watershed algorithm and overlaid on a SEM image of the sphere. (b) The average radius is calculated and plotted (black) against the actual perimeter (red). We find a standard deviation of 200 nm which amounts to a $\sim$0.7\% variation in the radius. 
	}
	\label{fig:Roundness}
\end{figure}

Concern about how variations in the sphere radius affect Casimir force measurements emerged because they were a possible, but unconfirmed, explanation for anomalous electrostatic force versus separation power laws \cite{Kim2008,Bezerra2011a}. 
Even though the expected electrostatic force power law is observed in many experiments \cite{DeMan2009,Chang2012}, concern about radius variations persisted because AFM measurements on spheres showed topographical irregularities \cite{Sedmik2013}. 
However, the tip shape of the scanning AFM probe imprints itself on the image, particularly when scanning the steep sides of the spheres \cite{Villarrubia1997}. 
Therefore, long-range deformation is instead calculated from SEM images. 
Here, we use an SEM (Hitachi S-3400) to assess the radius of our spheres. 
The sphere shown in Fig. \ref{fig:Roundness} is the sphere used for the Casimir force measurement presented in this manuscript.

To determine the roundness of the microspheres, we identify the circumference of the sphere through image segmentation via a watershed algorithm \cite{Sonka2007}
A centroid and boundary are extracted from the watershed result enabling the overlay of the identified perimeter of the sphere on the original SEM image as seen in Fig. \ref{fig:Roundness}a. 
Plotting the radius versus angle in Fig. \ref{fig:Roundness}b shows the ellipsoidal nature of the aberrations.
We attribute the ellipsoidal nature of the hollow glass microspheres to low energy ellipsoidal excitations while being formed from a liquid \cite{Lamb1916}. 
The standard deviation of the radius is then about 200 nm or 0.7\% of the radius, which minimally effects the electrostatic calibration (Fig. \ref{fig:SphereCapacitance}c). 
We note that an apparatus for measuring microsphere roundness more accurately and over the whole sphere, rather than just in profile, is currently under development \cite{Fan2014}.
\ntt{The radius calculated from an AFM scan, 33.2 $\mu$m, is within a micron of the radius measured with an SEM (32.54 $\mu$m), even though measurements of the sphere topography acquired using an AFM contain an imprint of the AFM probe}.

\subsubsection{Water layer}\label{s:waterlayer}

A thin water layer forms on most surfaces exposed to ambient conditions. 
The large relative DC permittivity of water ($\epsilon_{\nt{\text{W}}}$=77 at 303 K \cite{Kaatze1989}) causes even a nm-thick water layer to noticeably affect both the capacitance and the force.
As we are considering uncertainty in $d_{0}$ in this section, we focus on \nt{the effect of the water layer on} capacitance because it is used in the determination of $d_{0}$.
The effect of the water layer on the Casimir force itself is considered in a subsequent section.
The capacitance per unit area between two parallel conducting plates \nt{that are} a separation $d$ apart with a water layer of thickness\nt{,} $d_{\nt{\text{W}}}$\nt{,} on one of the surfaces is
\begin{align}
    \label{eq:waterlayer}
	C_{\text{pp}}(d_{\text{\nt{W}}})&=\frac{\epsilon_{0}}{d+d_{\text{\nt{W}}}(\frac{1}{\epsilon_{\nt{\text{W}}}}-1)}, \\
	&\approx\frac{\epsilon_{0}}{d-d_{\text{\nt{W}}}}, \notag
\end{align}
where $\epsilon_{\nt{W}}$ is the DC permittivity of water, and $C_{\text{pp}}$ is the capacitance between parallel plates.
Note that \nt{Eq. \ref{eq:waterlayer} implies that } uncertainty in $d_{\text{\nt{W}}}$ is roughly equivalent to uncertainty in $d_{0}$.
The relative increase of $C_{\text{pp}}$ due to the water layer is
\begin{align}
	W(d_{\text{\nt{W}}})=&\frac{C_{\text{pp}}(d_{\text{\nt{W}}})}{C_{\text{pp}}(d_{\text{\nt{W}}}=0)},
	\label{eq:waterprefactor}\\
    =&\frac{1}{1+\frac{d_{\text{\nt{W}}}}{d}\bigg(\frac{1}{\epsilon_{\nt{\text{W}}}}-1\bigg)}. \notag
\end{align}

Using the PFA to calculate the effect of water on the sphere-plate fit, we have
\begin{align}
	C'_{\text{PFA}} = 2 \pi R C_{\text{pp}},
\end{align}
so that when a water layer is included
\begin{align}
	C'_{\text{PFA}}(d_{\text{\nt{W}}}) =& 2 \pi R C_{\text{pp}}(d_{\text{\nt{W}}}), \\
    =&W(d_{\text{\nt{W}}})C'_{\text{PFA}}(d_{\text{\nt{W}}}=0).
\end{align}
Because the \nt{effect of the} water layer \nt{is the greatest over} the region of the sphere closest to the plate \nt{and vanishes at separations large compared to the water layer thickness}  (Eq. \ref{eq:waterprefactor}), the equation derived with the PFA is approximate when \nt{Eq. \ref{eq:full_cap} is} used for the capacitance
\begin{align}
	C'(d_{\text{\nt{W}}}) \approx W(d_{\text{\nt{W}}})C'(d_{\text{\nt{W}}}=0).
\end{align}
The effect of the water layer on $C''$ is similar:
\begin{equation}
	C''(d_{\text{\nt{W}}}) \approx W(d_{\text{\nt{W}}})^{2}C''(d_{\nt{\text{W}}}=0).
\end{equation}

Unfortunately, the thickness of the water layer can vary over the course of a measurement unless humidity is controlled precisely, and \nt{the water layer thickness itself} can vary across a single sample, particularly at grain boundaries. 
Moreover, estimates of $d_{\nt{\text{W}}}$ on gold vary widely depending on the type of measurement and the exact deposition process for the gold \cite{McCrackin1963,Gil2001,VanZwol2008b}. 
Without modeling or {\it in situ} measurement, the water leads to uncertainty in $d_{0}$ of about $\pm0.75$ nm in air for each surface (or 1.5 nm total). 
In addition, because the structure of water is much different at interfaces than in bulk \cite{Verdaguer2006,Kim2013a} it is plausible that a nm-thick layer of water affects $C'$ differently than would be expected from \nt{the} bulk properties \nt{of water}.
Finally, although the voltage applied between the sphere and the plate increases $d_{\nt{\text{W}}}$, calculations indicate that this change should be small compared to the overall water layer thickness \cite{Sacha2006a}.

\subsubsection{Roughness}\label{R1}

Roughness appears twice in the error analysis: first as uncertainty in the separation and calibration determination and second in the comparison of measurements to theory. 
Here we consider how roughness affect the separation uncertainty rather than how it changes the force on average. 

Many different roughness corrections have been developed for Casimir force measurements.
The first corrections were perturbative and assumed that the surface could be described by an average height with some standard deviation \cite{Boyer1994,Klimchitskaya1999a}.
Including the correlation length of the surface roughness leads to a correction that itself depends on the dielectric functions of the interacting materials \cite{Genet2003a}. 
After \nt{the} surface topograph\nt{ies of typical Casimir force probes were} found to follow a skewed probability distribution, new statistical methods were developed to account for the irregularity of the distribution \cite{Broer2012}.
Finally, the dependence on the particular orientation of the sphere, typically defined by the point of least separation (POLS) was noted, and a PFA-based technique was developed to estimate the uncertainty in the force from the uncertainty in the relative orientation of the sphere \cite{Sedmik2013}. 
The POLS is the point on the surface of the sphere that is also on the line between the center of the sphere and the closest point on the plate.

\begin{figure}[ht]
	\centering
	\includegraphics[width=.48\textwidth]{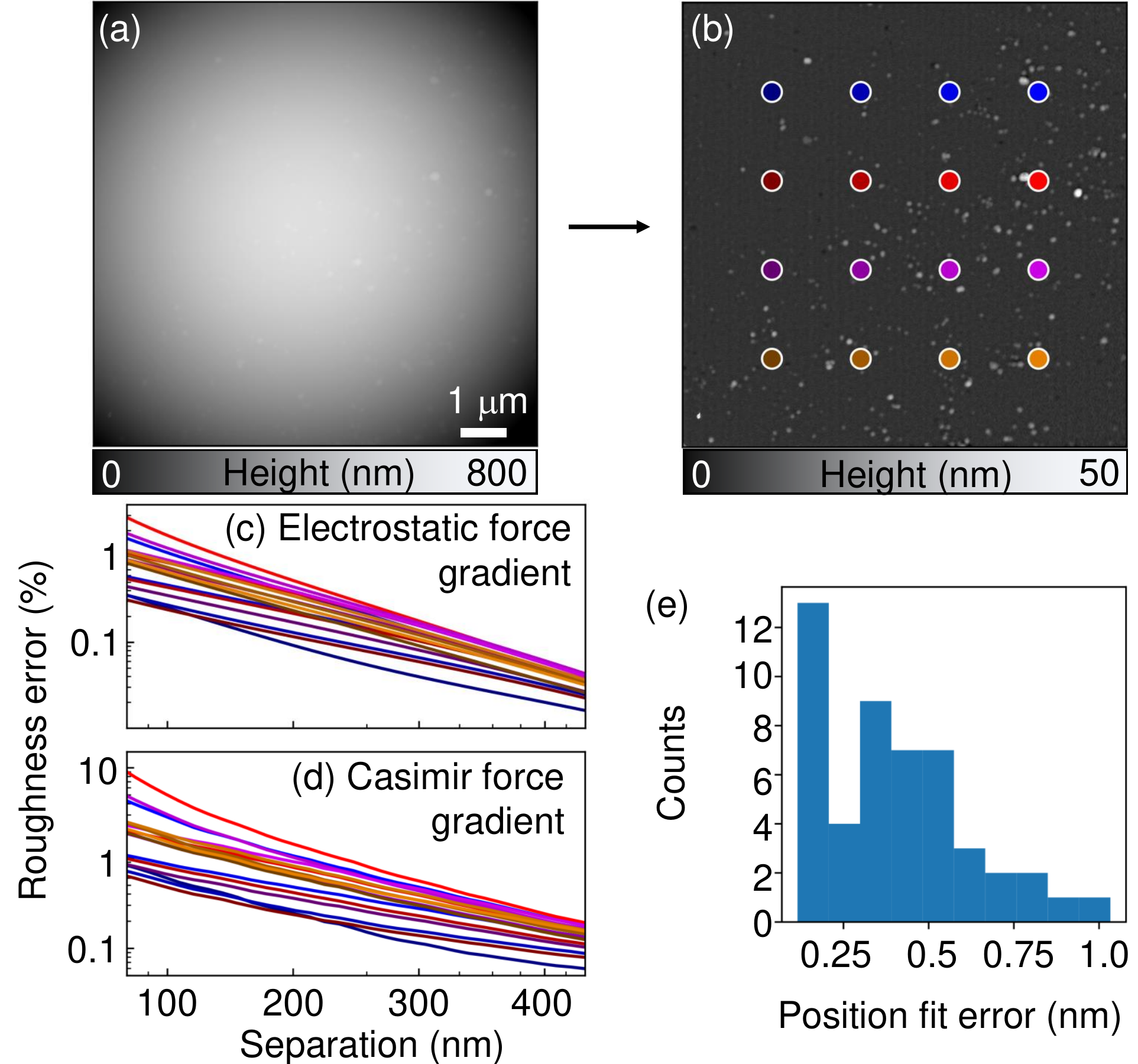}
	\caption[Calculating uncertainty from roughness using the proximity force approximation]{ 
		(a) An AFM image of the sphere is fit to the equation for a sphere and the fit is removed  (512 $\times$ 512 pixels). After the fit is removed, a 64 $\times$ 64 pixel median filter is used to separate the roughness from any imaging artifacts.
		(b) Several points are chosen on the roughness image to act as possible points of least separation (circles). 
		(c) The electrostatic force gradient for the sphere with roughness relative to a smooth sphere is calculated for each of the different points, identified by their color and brightness.
		(d) The Casimir force gradient for a rough sphere relative to a smooth sphere is also calculated for the nine points, and shows a much larger uncertainty because of the stronger separation-dependence of the force.
        (e) The roughness leads to a systematic offset in the separation determination up to 1 nm. 
	}
	\label{fig:roughness}
\end{figure}

Here, the uncertainty due to roughness is estimated using an oriented-PFA procedure akin to the one pioneered by Sedmik {\it et al.} \cite{Sedmik2013} (Fig. \ref{fig:roughness}). 
To prepare AFM topography scans of the spheres for a roughness analysis, the topography is first fit to the shape of sphere, $(x-x_{0})^{2}+(y-y_{0})^{2}+(z-z_{0})^{2}=R^{2}$, with the radius and center as free parameters. 
Then the fit is removed from the image. 
The resulting image retains systematic long-range distortions, from a combination of the imprint that the AFM tip leaves on the image \cite{Villarrubia1997} and imperfections in the sphere fabrication process.
To eliminate the distortions from the roughness analysis, the image is median filtered with a filter size larger than the short-range roughness (> 100 nm). 
The median-filtered image is then subtracted from the raw image so that only short-range roughness remains.

Because the sphere's orientation is only known to about 1 degree, the electrostatic force is calculated for many possible sphere orientations.
For each orientation, the measured topography is placed onto a model sphere of the appropriate radius.
The PFA is used to compute the roughness correction because roughness causes the largest effect at small separations. 
For the regions on the sphere where the topography is known, the force from the smooth sphere is subtracted and replaced with the force from the rough sphere pixel-by-pixel
\begin{align}
	\label{eq:Image2PFA}
	F_{\text{r}}(d) &= F_{\text{s}}(d) \\&- \sum_{i,j}\Bigg[F_{\text{pp}}(h_{\text{s}}(d,x_{i,j})) - F_{\text{pp}}(h_{\text{r}}(d,x_{i,j}))\Bigg],\notag
\end{align}
where $\sum_{i,j}$ is a sum over all the pixels in an image, $F_{\text{r}}$ and $F_{\text{s}}$ are the forces from the rough sphere and a smooth sphere, respectively, in the PFA limit, $F_{\text{pp}}$ is the force between each pixel and the pixel directly below it, and $h_{\text{s}}(d, x_{i,j})$ and $h_{\text{r}}(d, x_{i,j})$ are the separation between the surface of the sphere and the plate at that particular pixel for a smooth and rough sphere, respectively, when the point of closest approach is $d$ away from the plate. 
\nt{The electrostatic force on the rough surface is compared to the PFA calculation of the force on a smooth surface so that the same approximation is used throughout the calculation.}

The median rather than the mean is used to compute the height of the imaged portion relative to the smooth portion of the sphere because of the skewed height distribution.
Note that this formulation of a roughness PFA correction can be used to calculate either the electrostatic or Casimir force, as in \cite{Sedmik2013}, and works similarly well with force gradients.
Fig. \ref{fig:roughness}(c) shows that roughness primarily affects the electrostatic force near the surface. 
By fitting the electrostatic force gradient with the roughness corrections included, we determined that the uncertainty due to roughness on the sphere is $0.2$ nm (standard deviation of the offset from the 49 calculations, of which 16 are shown). 	

\subsubsection{Surface states}

The assumption that the macroscopic equation for capacitance is adequate for describing plate-plate and sphere-plate capacitance at the nanoscale has not been stringently tested for gold. 
For materials where this assumption has been tested (e.g. silicon and germanium), naively fitting a measured electrostatic force to the macroscopic form of the capacitance can lead to distance offsets between 60 and 600 nm, depending on preparation (for silicon), which were attributed to surface states \cite{Lamoreaux2012}. 

The offset for gold is likely less because it is more conductive and a recent measurement shows agreement between the predicted $C'$ and experiment out to 2 $\mu$m \cite{DeMan2009}, and here we observe $C''$ consistent with theory from 110 nm to 4 $\mu$m.
However, the presence of water or other adsorbates may complicate the nature of surface states. 
Further studies are necessary to determine the extent to which surface states affect $C'$.

\begin{figure}[ht]
	\centering
	\includegraphics[width=.4\textwidth]{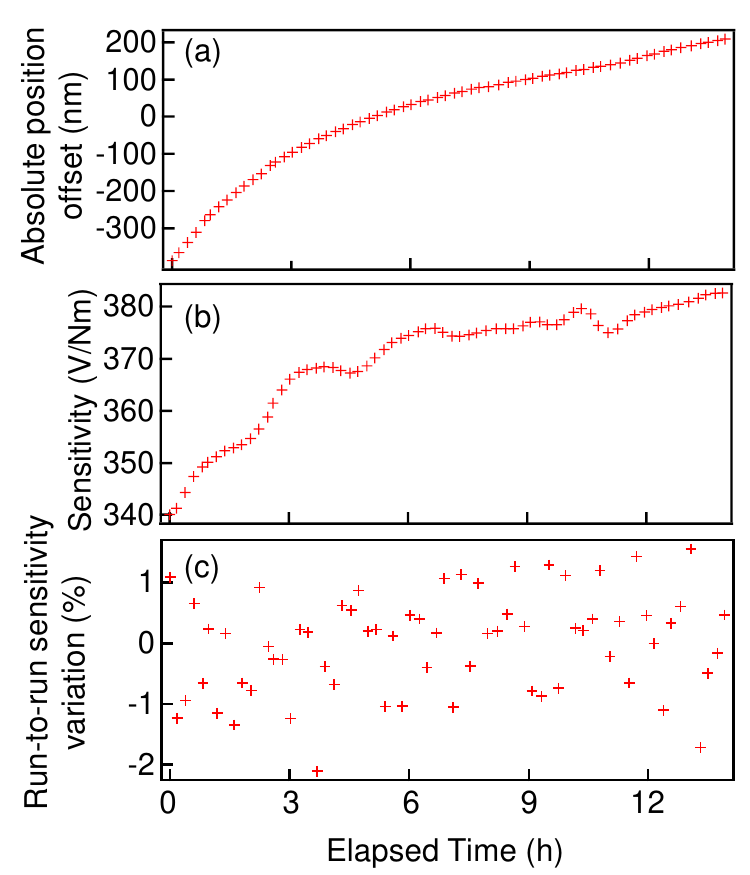}
	\caption[Variabtion in calibrations over a whole measurement]{ 
		(a) The position of the plate, $d_{0}$, drifts over time at a rate of about 50 nm/hr on average. For each run, a line is fit to $d_{0}$ versus time, including the two previous runs, the two subsequent runs, and the run itself. The drift is then deduced from the fit, and the linear drift correction is applied.
		(b) The force sensitivity increases by about 10\% over the measurement, but (c) the stochastic run-to-run variation is closer to $\pm$1\%. 
	}
	\label{fig:Stability}
\end{figure}	

\subsubsection{Drift}\label{drift}

Drift can both impart error to each individual determination of the sensitivity and separation and hinder the averaging of multiple data sets. 
To address drift in our experiments, the absolute separation, $d_{0}$, is determined for each individual approach or retract (Fig. \ref{fig:Stability}a).
The time dependence of $d_{0}$ is determined from the entire series of approach and retract runs and is found to drift at a rate of about 50 nm/hr, but decreasing as the elapsed time increases.
To account for the drift in the analysis of the measurement data, the drift within a run is determined by interpolating $d_{0}$ from the two runs before it and the two runs after it. 
By making $d_{0}$ itself a function of time, the effect of drift is accounted for in the data analysis procedure.

Likewise, drift in the electrostatic sensitivity calibration is also determined.
To understand the stochastic error and drift in the electrostatic calibration, the calculated absolute position offset $d_{0}$ and force gradient sensitivity are recorded for each approach (Fig. \ref{fig:Stability}).
Over the $13$ hours of measurements, the force gradient sensitivity shows systematic drift, and changes by a little over 10\%.
The change is likely due to drift between the photodiode and the photodetector used to track cantilever motion, so in the analysis the change is included in the optical lever sensitivity, $\gamma$, and its uncertainty.

\subsubsection{Cantilever bending}

\nt{Because t}he cantilever bends as it approaches the surface, \nt{the effect of the bending must be accounted for when determining $d$}.
The bending itself is often used for Casimir force measurements, but in our experiment deflection is detectable above the noise level only out to about 100 nm. 
The deflection signal ($\mu$V) is converted into a bending distance (nm) by dividing by the optical lever sensitivity $\gamma$. 
The static deflection is recorded at each height and a phenomenological power law is fit to the data to describe it because the deflection signal itself is too noisy to use as a correction on its own.
Then the recorded piezo extensions are adjusted to account for the cantilever's bending towards the surface based on the phenomenological fit.
Although bending leads to a small correction, it imparts uncertainty into the final separation, which is proportional to the uncertainty in the optical lever sensitivity $\gamma$.
If all the drift in the sensitivity is attributed to $\gamma$, then we can bound its variation by about $10\%$ over the measurement (Fig. \ref{fig:Stability}).
Because the bending is almost 3 nm at the closest approach, it adds about $0.3$ nm of uncertainty to $d_{0}$.

\subsubsection{Second-order oscillation}

The non-linearity of the electrostatic force not only leads to oscillations at harmonics, but also leads to higher-order corrections to the $S_{2\omega_{\text{A}}}$ signal. 
Because second-order oscillation is a dynamic effect it only appears in the $C'$ electrostatic measurement and not the $C''$ voltage parabolas, which incorporate only a static voltage. 
Using the PFA, the signal is \cite{DeMan2010}
\begin{align}
	S_{2\omega_{\text{A}}} &= -\frac{\gamma \epsilon_{0}\pi R V_{\text{AC}}^{2}}{2kd} - \frac{\gamma \epsilon^{2}_{0}\pi^{2}R^{2} V_{\text{AC}}^{4}}{2k^{2}d^{3}}-O(V_{\text{AC}}^{6}),\\
	&=-\frac{\gamma \epsilon_{0}\pi R V_{\text{AC}}^{2}}{2kd}\bigg(1 + \delta+ ...\bigg),\notag
	\label{eq:second_order}
\end{align}
where $\delta = \frac{\epsilon_{0}\pi R V_{\text{AC}}^{2}}{kd^{2}}$.
Now, estimating $\delta$ is equivalent to estimating the effect of the second-order oscillation on the separation determination. 
Because $V_{\text{AC}}$ is controlled by a feedback loop to produce a constant signal, $S_{\text{set}}$, during the measurement run, it is possible to solve for $\delta$, assuming it is much less than one:
\begin{align}
	\delta 
	&\approx \frac{2S_{\text{set}}}{\gamma d}.
\end{align}
With a typical cantilever (Table \ref{table:typical_cantilever}) and $S_{\text{set}}$ = 1 mV, $\delta\approx 0.014$ at 100 nm. 
Thus the correction is only a very small portion of the overall electrostatic signal.
However, the slow $d$-dependence makes it difficult to avoid the error without correcting for it. 
Note that these oscillations have a similar source as Eq. \ref{eq:4f_signal} and can, in principle, be estimated by measuring the 4$\omega_{\text{A}}$ signal.
However, recording 4$\omega_{\text{A}}$ with our AFM setup would require a four-step rather than a three-step measurement procedure, in order to collect all the data channels.
To estimate the effect of the second-order oscillation on the separation, we calculate how much the uncertainty in $\gamma$ leads to uncertainty in $\delta$. 
We then determine how much the uncertainty in $\delta$ affects the overall fit procedure by fitting with several different $\delta$ within the range of its uncertainty and predict that the imparted uncertainty is $0.3$ nm, but only in the $C'$-based separation determination because the $C''$-based determination uses only DC voltages. 

\subsubsection{Overall uncertainty in separation}

Most of the above sources of error tend to cause the surface to appear closer than it is. 
Moreover, these different sources of uncertainty can cause correlated error. 
The water layer thickness varies by a few nanometers in previous experiments depending on how it is measured \cite{McCrackin1963,Gil2001,VanZwol2008b}, so we posit a 1.5$\pm0.75$ nm water layer on each surface, which in turn leads to a $\pm$1.5 nm uncertainty in the separation of the two metal surfaces. 
Bending contributes about $\pm0.3$ nm of uncertainty, while roughness also contributes about $\pm0.2$ nm of separation uncertainty, so that the total uncertainty in position is about $\pm$2 nm for the $C''$-based separation determination.
For the $C'$-based method, there is an additional 0.3 nm of uncertainty from second-order oscillations. 

\subsection{Uncertainty in the measured signal}

\subsubsection{Systematic uncertainty in voltage offset}\label{sec:voltageuncertainty}

The first major source of uncertainty in $V_{0}$ comes from using $F_{a}$ to determine $V_{0}$ \nt{(Eq. \ref{eq:electrostaticforce})}.
$F_{a}$ has a much higher signal-to-noise ratio than $F_{DC}$, but it is also susceptible to AC coupling between the applied voltage and the drive piezo.
From K\nt{elvin} p\nt{robe} f\nt{orce} m\nt{icroscopy}, it is known that the voltage that minimizes the electrostatic force is not the same as the voltage that minimizes the electrostatic force derivative because the cantilever, rather than the tip, contributes much of the electrostatic force signal, but the tip contributes most of the force derivative signal \cite{Zerweck2005}.
For Casimir force measurements, the spherical probe has a much larger radius than an AFM probe (40 $\mu$m vs 30 nm), so the cantilever contributes a much smaller portion of electrostatic force.

The second source of uncertainty is present because an AC voltage is applied to the probe. 
The AC voltage applied to the probe can couple into the drive piezo, which leads to an extra signal fed into the voltage feedback loop \cite{Melin2011,Barbet2014}. 
The additional signal combined with the separation-dependence of the electrostatic force leads to a distant-dependent artifact in $V_{0}$. 
Any generic offset of the output of a lock-in amplifier, in fact, leads to such an error. 
The voltage artifact is proportional to 1/$C'$, and knowledge of $C'$ permits an estimate of the voltage error. 
If all the separation-dependence of  $V_{0}$ is attributed to the extraneous voltage, then estimates can be made of the original offset and the residual electrostatic force that remains because of the extraneous voltage. 
Based on the separation-dependence of $V_{0}$, we estimate that the offset in the signal is less than 10 $\mu$V, which would lead to an error in $V_{0}$ of less than 10 mV at separations where the Casimir force is measured.

\subsubsection{Optical interference}\label{sec:interference}

Because the optical lever used to detect the motion of the cantilever is coherent, an interference pattern appears in the response signal of the cantilever to the shaking plate. 
A small amount of the optical beam that falls off of the cantilever ({\it e.g.} Airy disks) has a different path length to the detector than the light reflected directly off the cantilever \cite{Huang2006}. 
As the surface is brought towards the cantilever, the interference condition at the photodiode changes. 
The interference artifact is common to AFMs that use optical lever detection \cite{Huang2006} and has been identified before in Casimir force measurements as a factor that limits accuracy at large separations \cite{DeMan2010,Sedighi2016}.	

\begin{figure}[ht]
	\centering
	\includegraphics[width=.49\textwidth]{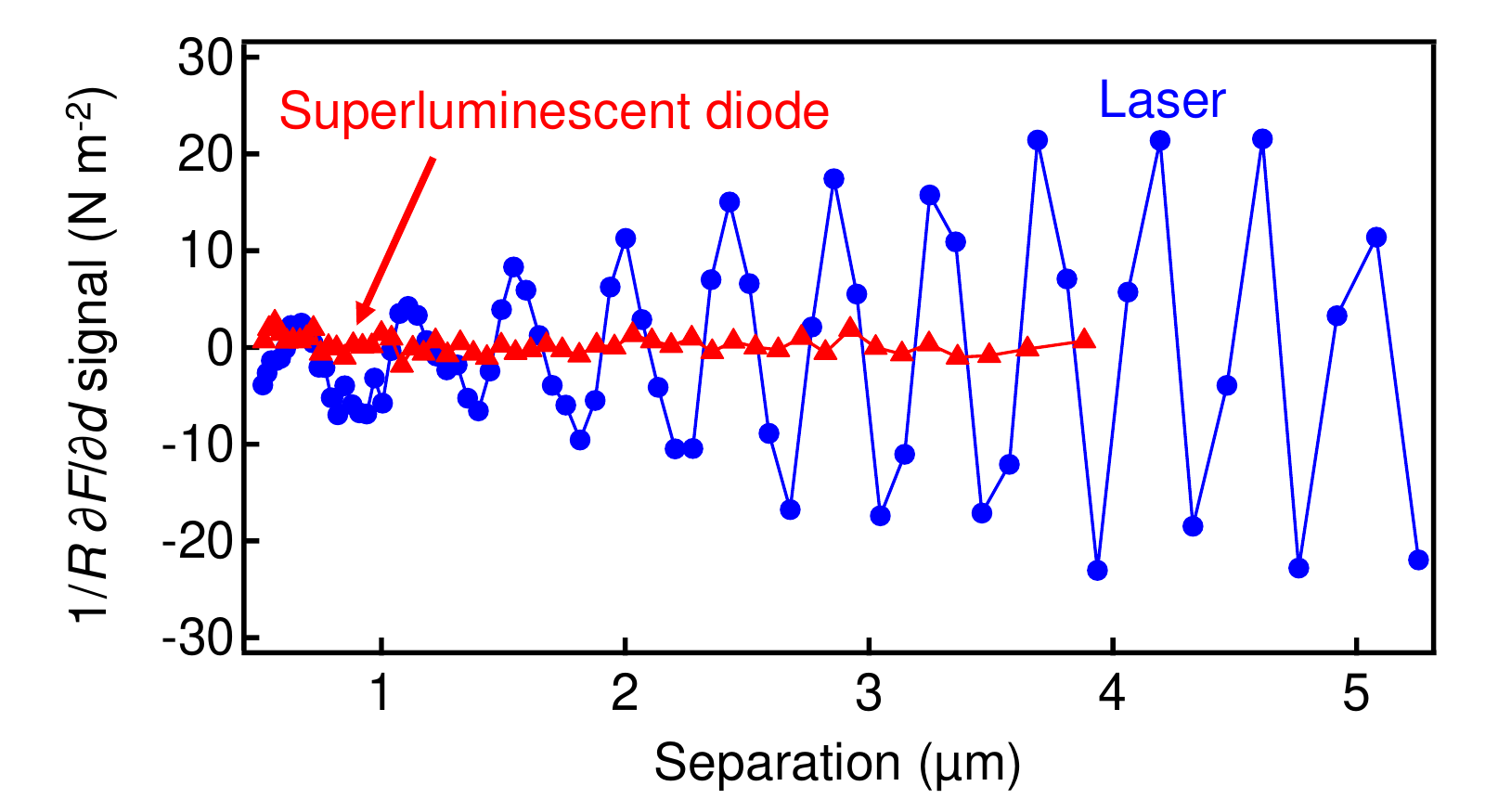}
	\caption[Comparing laser and superluminescent diode interference]{ 
		The interference artifact is analyzed by comparing the calibrated force gradient signal at distances larger than 500 nm. The laser causes interference fringes on the order of $\pm10$ N m$^{-2}$ (red), but the superluminescent diode (SLD) produces fringes with a magnitude of about $1$ N m$^{-2}$ (for a 40 $\mu$m radius sphere). Thus, the SLD detector reduces the error imparted by interference by a factor of 10. Because the interference appears in the force data and is linear in shake amplitude like the force itself, it is reported, after calibration, in N m$^{-2}$.
	}
	\label{fig:Interference}
\end{figure}

In order to minimize the optical interference in the Casimir measurements presented here, two different optical sources are tested. 
A diode laser and a superluminescent diode (SLD) \cite{Alphonse1988}, which limits the coherence of the light by increasing its wavelength spread, are tested using the same force measurement procedure ($\lambda=$ 860 nm for both).
The interference appears in the force data channel. 
Figure \ref{fig:Interference} shows the signal for the approaching cantilever once with the laser and once with the SLD. 
The SLD is confirmed to decrease the interference artifact in the force signal by about an order of magnitude relative to the laser.  
Even so, the interference from the SLD appears in the data at a level of about 1 N m$^{-2}$. 

The force uncertainty from interference is estimated by fitting the data from the \nt{force gradient signal channel} to sine waves at separations > 500 nm. 
The \nt{least-squares} fit \nt{to the sum of a sine curve at} the primary wavelength in the interference, which is half the wavelength of the source (\nt{$\lambda/2$}) \nt{and a sine curve at} wavelength of the next harmonic ($\lambda/4$). 
Only the amplitude and phase of one of wavelengths is permitted to vary at a time in the fit procedure. 
The amplitudes of both \nt{sine curves} are summed for a rough estimate of the uncertainty imparted by interference. 
Even though the fits characterize the uncertainty, attempts to use the fits to remove the interference after measurement \nt{are} unsuccessful.
This is because the two spatial frequencies do not completely describe the interference and the interference may change its amplitude at at different separations, as it does between 2 and 4 $\mu$m in Fig. \ref{fig:Interference}. 
The interference artifact varies by \nt{up to} a factor of 4 between measurements \nt{when the cantilever probe varies but the light source remains the same}.

\subsubsection{\nt{T}he hydrodynamic force}\label{sec:hydrodynamic}

For dynamic measurements in air, the hydrodynamic force, $F_{\text{H}}$, is of comparable magnitude to the Casimir force.
The Casimir force decays more rapidly with increasing separation, so that the hydrodynamic force limits how far from the surface it can be observed. 
However, the hydrodynamic force is proportional to velocity, which is 90 degrees out of phase with the displacement of the plate. 
A lock-in amplifier separates the in-phase from the quadrature signal in order to separate the hydrodynamic force from the Casimir force (Fig. \ref{fig:MeasurementTechnique}).

\begin{figure}[ht]
	\centering
	\includegraphics[width=.4\textwidth]{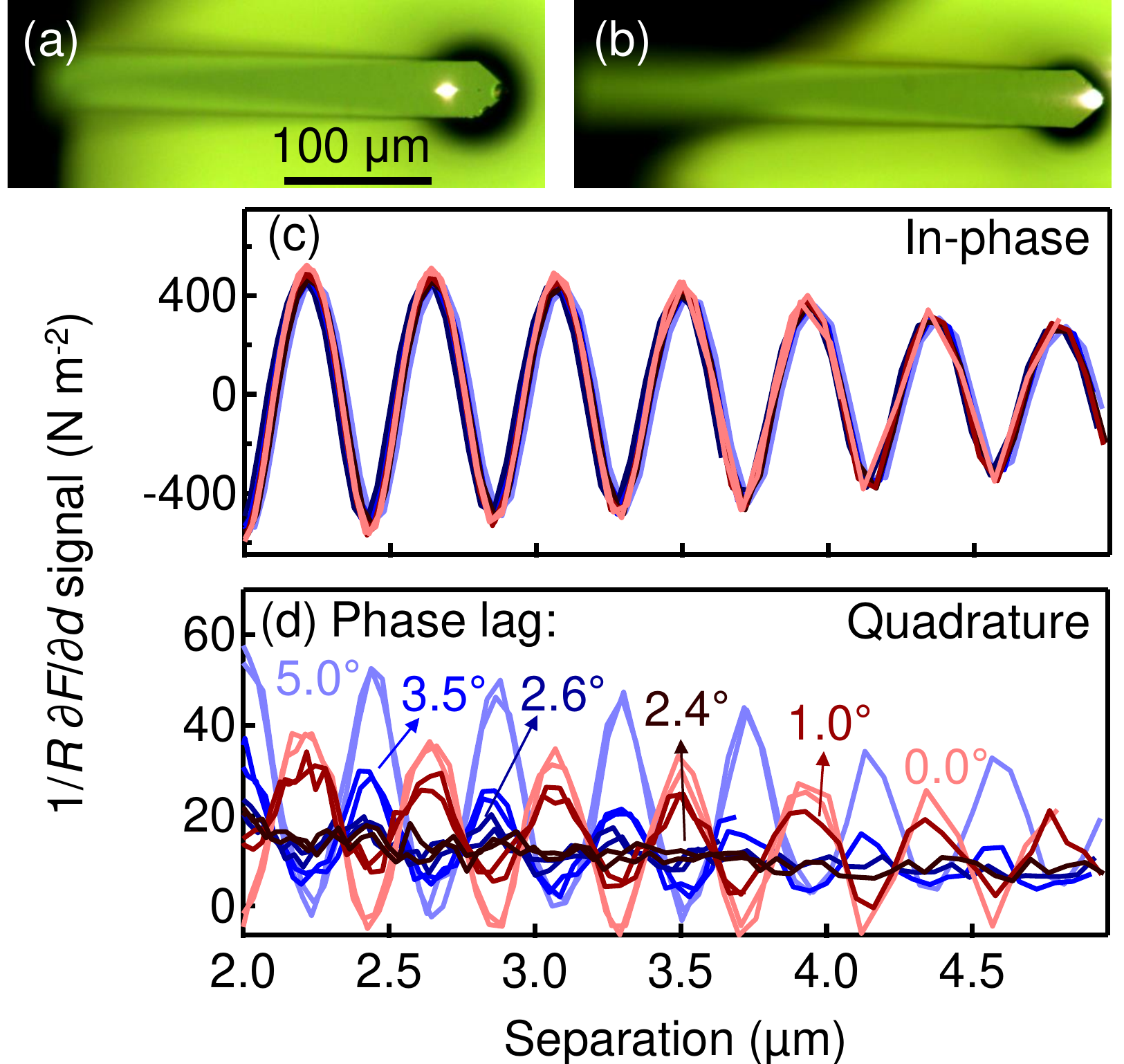}
	\caption[Setting phase with interference]{ 
		(a) The detection beam is focused at least 15 $\mu$m away from the edges of the cantilever, in order to minimize interference during force measurements. 
		(b) The interference is greatly increased by focusing the detection beam at the tip of the cantilever.
		(c,\nt{d}) Interference is used to determine the reference phase of the lock-in amplifier to within about $0.2$ degrees because, when the reference phase is set properly, the interference only appears in the in-phase channel of the lock-in amplifier. 
		The hydrodynamic force can be excluded from the Casimir force channel more completely (0-5 degree reference phases shown).  
        For these measurements, a phase lag of 2.4 degrees is optimal. 
	}
	\label{fig:Settingphase}
\end{figure}

\nt{The accuracy of the reference phase, $\theta_{\text{ref}}$, determines the imprint of the hydrodynamic force in the Casimir force measurement signal.
Uncertainty in the phase has two parts: a constant phase offset and a phase offset that depends on separation, due to dissipation.}
The delay between the direct digital synthesizer and plate must be measured in order to set the reference phase to sufficient accuracy, because the hydrodynamic force enters into the Casimir force signal as $F_{\text{H}}\sin(\Delta\theta_{\text{ref}})$, where $\Delta\theta_{\text{ref}}$ is the error in the reference phase, which is about 0.2 degrees in our measurement.
The uncertainty from the hydrodynamic force is calculated from the hydrodynamic force measured by the quadrature channel multiplied by $\sin(\Delta\theta_{\text{ref}})$.

While interference is problematic in the force measurement, it can be utilized to set the reference phase of the LIA that records the response of the cantilever to the shaking plate (Fig. \ref{fig:Settingphase}). 
Because the interference is determined by the position of the plate, relative to the cantilever, and is independent of the velocity of the plate, it appears in the in-phase channel and is excluded from the quadrature channel of the LIA. 
To determine the reference phase, we first replace the SLD with the laser light source and focus it at the edge of the cantilever in order to accentuate the amount of interference.
Second, the cantilever approaches and retracts from the surface with several different reference phases. 
The reference phase for which the interference falls entirely in the in-phase channel of the lock-in amplifier is chosen for use in force measurements. 

\begin{figure}[ht]
	\centering
	\includegraphics[width=.49\textwidth]{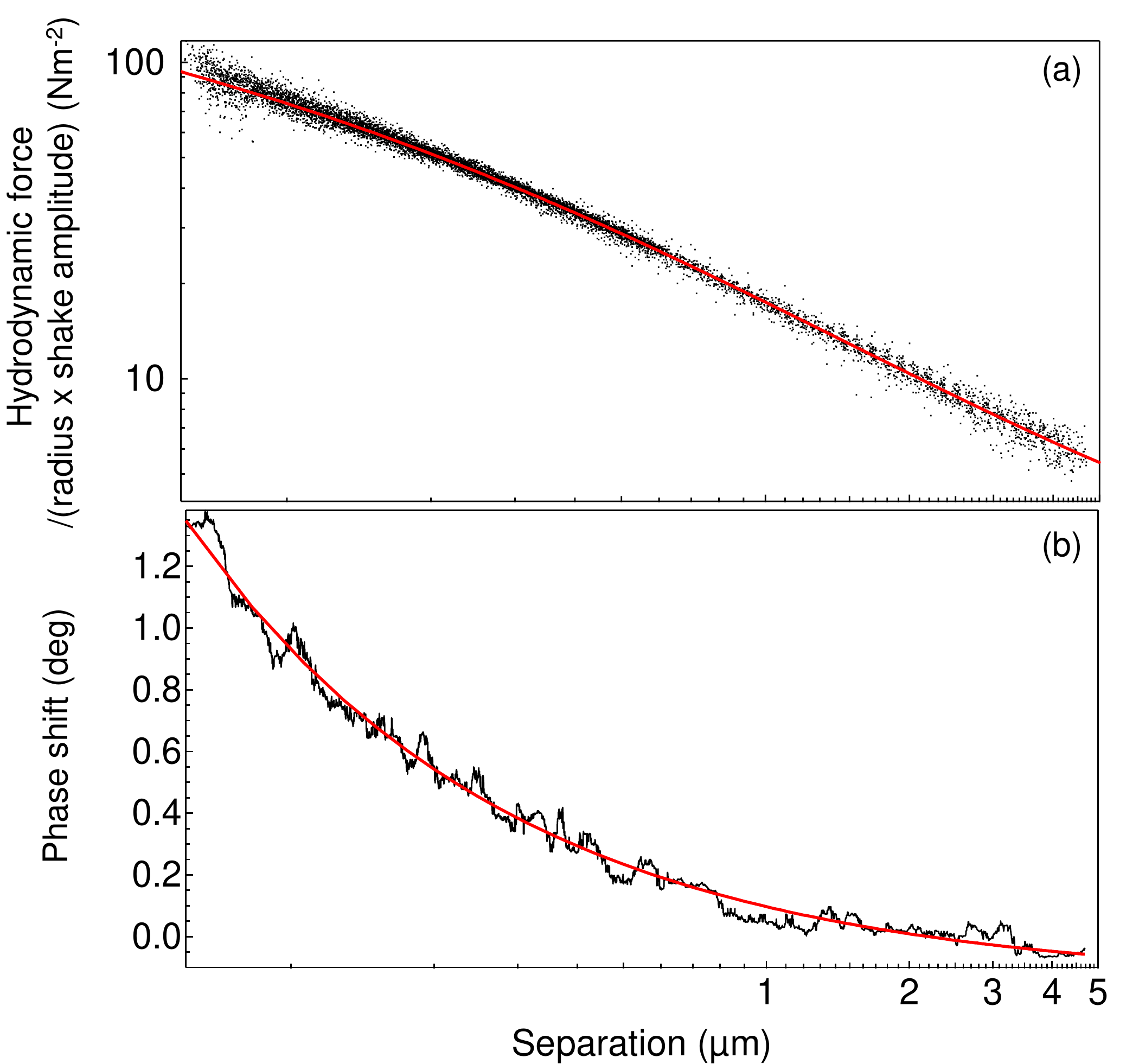}
	\caption[phase figure]{ 
		\nt{The hydrodynamic force (a) is measured as well. The increase in the hydrodynamic force increases the phase shift of the cantilever's response (b). The phase shift caused by the hydrodynamic force is measured by monitoring the phase of the electrostatic $S_{2\omega_{\text{A}}}$. The red lines are fits used in the analysis of the uncertainty coming from the hydrodynamic force.} 
	}
	\label{fig:phasemeasurement}
\end{figure}

The accuracy of the interference method of setting the reference phase is \nt{also} limited by changes to the cantilever's transfer function due to the hydrodynamic force.
The phase lag is $[Q(d)]^{-1} (\omega/\omega_{1})$ in the $\omega \ll \omega_{1}$ limit, where $Q$ is determined by the hydrodynamic damping \cite{Vinogradova1995,Maali2008,Laurent2011a}
\begin{align}
	Q &= \frac{k}{\omega_{1}}\bigg(\Gamma_{0} + \frac{6\pi\eta R^{2}}{d}f^{*}(d/6b)\bigg)^{-1},
\end{align}
where $\Gamma_{0}$ is the damping of the probe far from the plate, $\eta$ is the dynamic viscosity, $b$ is the slip length, and $f^{*}$, called a correction function in \cite{Laurent2011a}, is a monotonic function that approaches 1 for $d \gg b$ and approaches 0 for $d \ll b$. 
\nt{The phase lag of the cantilever's response is measured from the frequency shift of the $S_{2\omega}$ electrostatic signal (Fig \ref{fig:phasemeasurement}).
Once the phase lag is known as a function of separation, it is incorporated into the force measurement and uncertainty analysis.
First, the expected phase shift of the cantilever response at the piezo shake frequency is calculated as $\phi_{c}(\omega_{\text{pz}}, d)\approx [\omega_{\text{pz}}/(2\omega_{\text{A}})]\phi_{c}(2\omega_{\text{A}}, d)$. 
Second, $F_{\text{H}}\sin(\phi_{c}/2)$ is subtracted from the measured force gradient signal. 
Third, $|\phi_{c}/2|$ is added to the reference phase uncertainty. 
The uncertainty is consistent with the electrostatic estimate of the phase lag and the estimate of no phase lag.}

\section{Total measurement uncertainty}\label{s:TotalUncertainty}

\begin{figure}[ht]
	\centering
	\includegraphics[width=.48\textwidth]{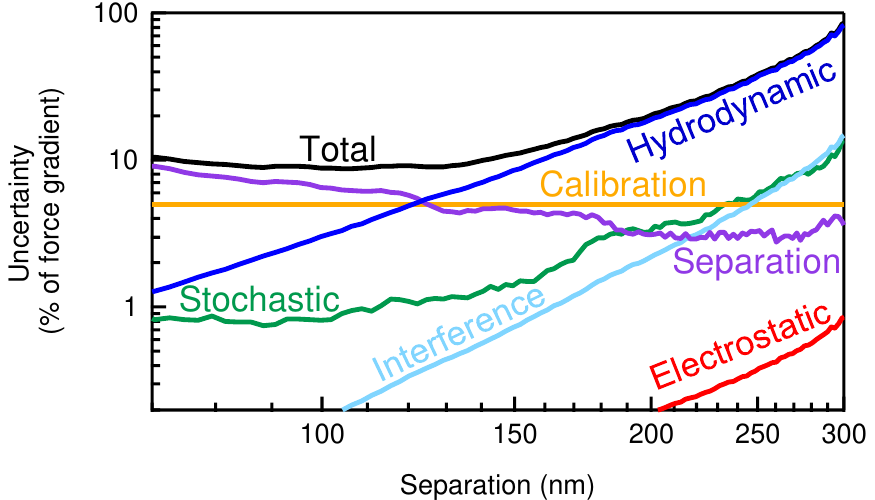}
	\caption[Uncertainty in force measurements]{ 
		The expected uncertainty in the Casimir force measurements is calculated from several sources of error. 
        At short range, separation determination (section \ref{sec:separation}) dominates the error.
        Although the force is detected well above the noise level there, uncertainty in the position makes experiment-theory comparison less clear. 
        At large separations the hydrodynamic force, \nt{as well as} interference and stochastic noise, dominate (sections \ref{sec:interference}, \ref{sec:hydrodynamic}, and \ref{sec:thermal}).
	}
	\label{fig:AllErrors}
\end{figure}

To understand how the different sources of error contribute to the force measurement at different separations, the uncertainties are combined at separations from 30 nm to 300 nm (Fig. \ref{fig:AllErrors}).
They are added in quadrature, under the assumption that each source is uncorrelated with the others.
At short separations, separation uncertainty is the dominant contribution.
At large separations, interference, stochastic noise, and the hydrodynamic force dominate the uncertainty. 
The force sensitivity is limited to about 2 pN when approximated from the smallest observable force.
Therefore, significant reductions in uncertainty are possible. 

The measured force sensitivity compares favorably to other AFM force measurements, which report an optimal sensitivity of $\sim1$ pN \cite{Butt2005}. 
Other Casimir force measurements in the sphere-plate geometry report a force sensitivity of 2-5 pN in air \cite{VanZwol2008c,DeMan2010}, while in the parallel plate geometry the sensitivity is limited to a few nN (but with a much larger interaction area) \cite{Almasi2015a}. 
A few measurements of the Casimir force in vacuum report force sensitivities at the fN level with a $\sim$1 s integration time per separation \cite{Decca2005,Chang2012} (using equivalent noise for force gradient measurements \cite{Kobayashi2009}). 
Sensitivities at or below the fN level are reported in fluid \cite{Ether2015,Liu2016}, but because the spheres used were all <  10 $\mu$m and the electric double-layer force was present, the Casimir force was not unambiguously observed. 
Measurements in liquid that do observe the Casimir force report a 50-100 pN sensitivity \cite{Munday2007,VanZwol2009}. 

Comparing Casimir force measurements performed with an AFM to other microsphere measurement technologies, such as torsion pendulums and optical traps, shows that AFM measurements have lower overall sensitivity but offer more control.
In vacuum, force sensitivities using microspheres in optical traps report measurements up to aN level sensitivity \cite{Ranjit2016a,Blakemore2018} but, though there has been much recent progress, controlling the position relative to a surface remains a challenge.
Torsion pendulums give a pN level sensitivity with much larger spheres ($R\approx$ 1 cm \cite{Lamoreaux1997,Lamoreaux2005}).
AFM measurements, such as those presented here, allow control over orientation, which enables measurements between surfaces of different shapes and characterization of the exact regions of the surfaces that are interacting \cite{Garrett2018}.
Even though optical traps and torsion pendulums provide greater sensitivity or interaction area, AFM probes will continue to have a role in Casimir force measurements because they allow the investigation of intricate geometric orientations and the detailed characterization of the interacting surfaces. 

\subsection{Reducing measurement uncertainty}

Based on the above analysis, there are two routes to reduce the uncertainty in Casimir force measurements. 
Near the surface, improvements in the separation determination reduce uncertainty the most.
Far from the surface, improvements to interference, stochastic noise, and the hydrodynamic force reference phase all reduce the total uncertainty. 
Reductions to calibration uncertainty reduce uncertainty everywhere.
Investigations into the reliability of calibrations using the electrostatic calibration could be performed by comparing them to calibrations performed with optical or other forces \cite{Weld2006a,Liu2016,Maali2008}. 
A better understanding of calibration uncertainty is critical to Casimir force measurements, because it affects them at all separations.
Because of similarities between the measurement presented here and prior measurements, we expect that the following uncertainty reduction strategies will also lower error in other Casimir force measurements as well.

\subsubsection{Near the surface}

At small separations, the Casimir force can be measured well above the stochastic noise level. 
To improve the measurement, it is necessary to improve the separation determination.		
Because many factors contribute to the uncertainty in the separation as determined by the electrostatic force, it would be infeasible to address them simultaneously. 
Some, such as the presence of a water layer, could be addressed with improved characterization of the samples.
Most inhibitive is the presence of surface states which would require significant experimentation in surface science to characterize.
Therefore, one tactic to evaluate and improve the accuracy of the separation determination is to develop new ways to measuring separation and comparing them \cite{Chang2004}. 

A direct way to measure the position of the surface is through contact measurements, but roughness adds significant uncertainty in the relationship between the distance-upon-contact and the absolute separation of the two surfaces. 
However, for sufficiently smooth surfaces, the difference vanishes. 
The electrostatic force is typically used to determine the absolute separation because it has a strong separation-dependence that is described by an analytic formula.
In addition, the hydrodynamic force has been used successfully for separation determination in liquids \cite{Munday2007}.
It might also work in air even though the slip length is considerably larger ($\sim$60 nm versus < 10 nm) because it can be amplified by using larger probes and higher frequencies \cite{Maali2008a,Laurent2011a}.
Using the hydrodynamic force in air may also permit Casimir force measurements with insulators, as is possible in liquid \cite{Munday2009a}.

\subsubsection{Far from the surface}

The Casimir force is predicted to be observable out to a separation about four times larger than reported in this manuscript. 
Therefore, at large separations, there is potentially more to reveal about the Casimir force by decreasing the uncertainty. 
The hydrodynamic force can be made smaller by shaking the plate at a lower frequency, by varying the reference phase with separation, or by using smaller spheres \nt{because the hydrodynamic force is proportional to $R^{2}$ rather than $R$ \cite{Maali2008}}. 
The interference is harder to eliminate because the SLD is already designed to minimize coherence.

One possible way to minimize the interference would be to measure the position of the cantilever with light at several wavelengths. 
Other possibilities include measuring the force with the optical lever at a few different positions along the back of the cantilever to change the path length of the interference, or varying the focus of the light onto the cantilever. 
Different detection techniques, such as laser doppler vibrometry \cite{Labuda2015} or piezoelectric cantilevers \cite{Watanabe1996}, might circumvent the artifact. 
Uncertainty from stochastic noise can be reduced by taking more data, using a larger shake amplitude, operating at a lower temperature\nt{, or finding a lower-noise photo-detector}.

\section{Comparison to other techniques}

\nt{Above, several major sources of error in force modulation measurements have been listed and quantified as a function of tip-sample separation, and the uncertainty in the Casimir force measurements is calculated from them. 
In this section, the uncertainties of the force modulation method are compared to those in the frequency modulation and deflection methods.}

\subsection{Frequency modulation}

Measurements using frequency modulation (FM) detection with AFM probes in vacuum report the highest precision of any Casimir force measurements \cite{Decca2005,Chang2012}. 
In the FM detection scheme, the force gradient is measured through the change of the first resonance frequency of the cantilever as the sphere approaches the surface. 
The frequency shift is given by
\begin{align}
\Delta \omega_{1} \approx -\frac{\omega_{1,0}}{2 k}\frac{\partial F}{\partial d},
\end{align}
where $\omega_{1,0}$ is the resonance frequency of the first eigenmode far from the surface and $\Delta \omega_{1}$ is the change in the resonance frequency at separation $d$. 
Note that the resonant frequency is then $\omega_{1}$ = $\omega_{1,0}+\Delta\omega_{1}$.
To date, all reported Casimir force experiments using the FM technique took place in vacuum environments. 
Our attempts toward an FM measurement in air and the artifacts we found in that environment are \nt{discussed in \cite{Garrett2017d}}.

\nt{Frequency modulation measurements, despite their higher overall sensitivity, are subject to several additional sources of error in air. 
First, direct piezo-actuation of the cantilever, at frequencies above the first resonance of the piezo, hides the cantilever resonance in a `forest of peaks' \cite{Kaggwa2008,Proksch2010,Labuda2011a, Labuda2011}, which increases the uncertainty in the determination of the resonance frequency. 
Second, as the cantilever approaches the surface and is damped by the hydrodynamic force, the error in the resonance frequency grows in proportion to the damping, which leads to an artifact proportional to $d^{-1}$. 
Third, any coupling between the voltage applied to the drive piezo, used to maintain constant excitation amplitude, and the lock-in amplifier causes an artifact proportional to $d^{-2}$. 
In the experiments discussed in \cite{Garrett2017d}, these three artifacts prevent us from observing the Casimir force in air using the frequency modulation method.}

\subsection{Deflection}

Several experiments in air have measured the Casimir force through the detection of the cantilever's deflection \cite{VanZwol2008c,Sedighi2016}.
At any one height, the deflection is $\mathfrak{D}=F(d)/k$, but low-frequency $1/f$ noise dominates the signal, which leads to a trade-off: acquiring force curves faster excludes more low-frequency drift, but also leads to more correlation between the error at nearby separations.
Moreover, increasing the speed at which the data are collected causes the hydrodynamic force, which is proportional to velocity, to be present in the data at higher levels.
In addition, repeated contact with the surface during measurements can damage the tip. 
While damage does not always occur and can be observed after the measurement by AFM or SEM images of the probe itself, it can be difficult to identify when during a set of measurements the probe is damaged.

\section{Uncertainty from the force calculation}

\begin{figure}[ht]
	\centering
	\includegraphics[width=.45\textwidth]{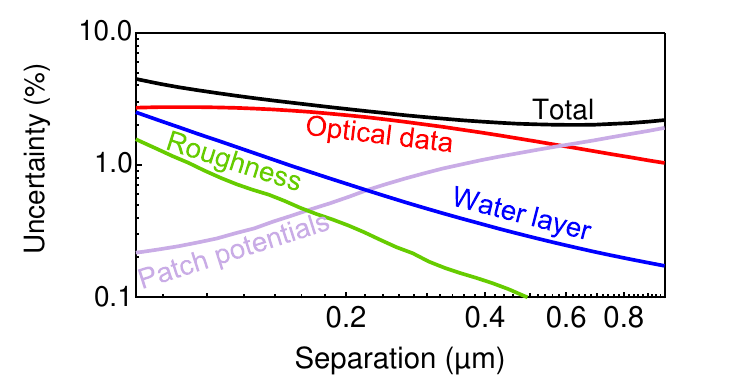}
	\caption[Uncertainty in force calculation]{ 
		The uncertainty in the force calculation comes from uncertainty in the dielectric constant, water layer thickness, roughness and patch potentials.
	}
	\label{fig:Funcertainty}
\end{figure}

The uncertainty in Casimir force data comes not only from the measurement error, but also from uncertainty about the sample being used, which includes uncertainty regarding optical properties \cite{Pirozhenko2006,Munday2008c,Svetovoy2008}, roughness \cite{Genet2003a,VanZwol2008c,Broer2012,Sedmik2013}, and patch potentials \cite{Speake2003,Kim2010,Behunin2012,Behunin2014a,Garrett2015}.
Because of these factors, the calculated total force has uncertainty itself. 
Of the different uncertainties, the uncertainty in the gold's optical properties is the limiting uncertainty over most of the range (Fig. \ref{fig:optical_properties}).
At the shortest separations, the water layer and roughness become larger sources of uncertainty (Fig. \ref{fig:Funcertainty}). 

\subsection{Sample dielectric function}\label{s:optical}

\begin{figure}[ht]
	\centering
	\includegraphics[width=.45\textwidth]{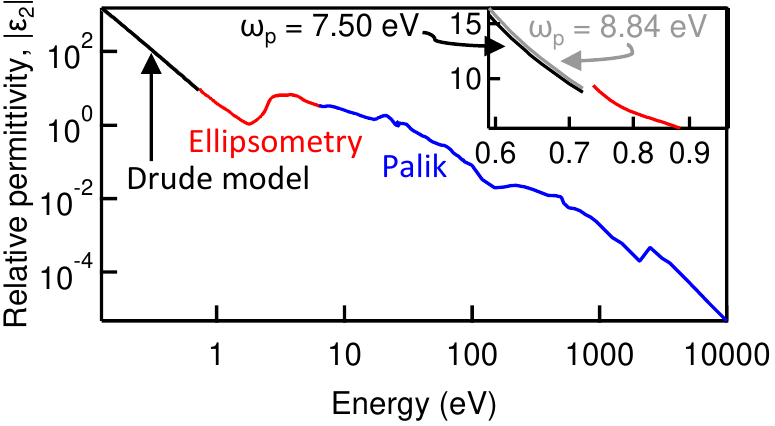}
	\caption[Dielectric function of gold sample]{ 
		The dielectric data used to estimate the Casimir force is computed from ellipsometry data in the range of 0.73-6.3 eV combined with Palik reference data at higher energies and the Drude model at lower energies. The inset shows the boundary between the measured data and different fits for the Drude model.
	}
	\label{fig:optical_properties}
\end{figure}

Uncertainty in the dielectric function of the interacting surfaces leads to uncertainty in the calculated Casimir force. 
Because $\epsilon_{\text{air}}\approx1$, the two gold surfaces contribute most of the uncertainty to the Casimir force measurement. 
Because tabulated optical data used on its own leads to 5-15\% uncertainty in the force \cite{Pirozhenko2006,Munday2008c}, the dielectric response is measured with ellipsometry of an evaporated 100 nm Au film on a glass slide in the 0.73 to 6.3 eV range (Fig. \ref{fig:optical_properties}). 
The ellipsometry data are then compared to the tabulated Palik data \cite{Palik1998}. 
Because of the agreement with the ellipsometry data at high energies, the Palik data at energies above those collected with ellipsometry are used. 
The tabulated dielectric data agree with the measurement less well at low energies, so the response there is extended with the Drude model. 
Pirozhenko {\it et al.} \cite{Pirozhenko2006} lists the Drude model parameters for several different samples of gold. 

By comparing measured ellipsometry data to the Drude parameters, the data from ellipsometry are determined to be most similar to a plasma frequency $\omega_{\text{p}} = 8.84$ eV and $\omega_{\tau} = 0.042$ eV or $\omega_{\text{p}}=7.50$ eV and $\omega_{\tau}=0.061$ eV (Fig. \ref{fig:optical_properties}).

Because the resulting force difference from uncertainties in modeling the Drude parameters is larger than the difference between the force when calculated with either the plasma or Drude model with the same plasma frequency ($\sim 1 \%$) \cite{Chang2012}, the experiments presented here are not yet at the level of accuracy to be able to comment on that discrepancy.
The force is computed using the combined optical data together with each set of reference Drude parameters.
The difference between the two calculations is used as the uncertainty from the optical properties \cite{Lambrecht2006a}.

\subsection{Patch potentials}

The force from patch potentials on gold tends to be about 1\% of the Casimir force over the pertinent measurement range, but it has become a major concern in Casimir force experiments because it tends to follow a similar separation-dependence. 
Because the patch potential contribution to the total force cannot be separated from the Casimir force during an experiment, we include it as a source of uncertainty in the calculation of the total expected force.

A few experiments used Kelvin probe force microscopy images of the surface potential on a plate to calculate the patch potential force between a sphere and a plate, using either the assumption that the potentials on a sphere and a plate are statistically identical \cite{Behunin2014a,Garrett2015} or by comparing samples with different patches \cite{Garcia-Sanchez2012a}.
For the estimation of uncertainty presented here, the calculated patch potential forces from \cite{Garrett2015} are used.
Note that the uncertainty comes from the sample-to-sample variation in the patch potential force, rather than its average value.

\subsection{Calculating the Casimir force with roughness}\label{sec:FGroughness}

Roughness also adds uncertainty to the calculated force. 
Atomic force microscopy is used to measure the roughness on both the sphere and the plate, as has been performed before \cite{Klimchitskaya1999a,VanZwol2008c,Sedmik2013}.
If the relative positions of sphere and plate are known, then the predicted forces can be calculated directly from the topography images \cite{Lambrecht2006a,Rodriguez2007,Woods2016,Hartmann2017}. 
However, there is uncertainty in the exact orientation of the sphere because the point of closest approach is known only to a few microns and the exact position above the plate is unknown as well. 

The spheres tend to be much rougher than the plates because the fabrication processes for hollow spheres have been developed only recently, and, while precision fabrication techniques exist \cite{Campbell1983}, our spheres were procured from a commercial source (Trelleborg SI-100).
Because the sphere tends to be much rougher than the plate, the focus of the roughness uncertainty comes from uncertainty in the orientation of the sphere
 \cite{Sedmik2013}. 
The technique used to calculate electrostatic roughness corrections is used again  (Fig. \ref{fig:roughness}d).
To compute the roughness uncertainty, the Casimir force gradient is calculated for 49 different points on the sphere profile, and the uncertainty is computed as the range around the most likely estimate within which about 68\% of the calculated roughness corrections fall. 
Note that the distribution of corrections is extremely irregular (Fig. \ref{fig:roughness}). 

\subsection{Calculating the Casimir force with a water layer}

The water layer discussed above affects not only the separation determination, but also the calculation of the Casimir force itself. 
Because the presence of a water layer on the metal surfaces tends to increase the Casimir force between metal plates, due to a decrease in the absolute surface separation \cite{Palasantzas2009}, uncertainty in the thickness of the water layer leads to uncertainty in the Casimir force theory. 

To investigate the effect of the water layer thickness uncertainty, we calculate the Casimir force with a 0.75, 1.50, and 2.25 nm thick water layer on each surface. 
The uncertainty of the water layer is calculated as the average of the differences of the 1.50 nm calculation with the 0.75 and 2.25 nm calculations at each separation (e.g. uncertainty = $|F_{0.75}-F_{1.50}|/2+|F_{2.25}-F_{1.50}|/2$). 
At small separations, the water layer becomes the largest source of error in the calculation. 

\section{Conclusions}
A measurement of the Casimir force has been presented, as well as several experiments designed to characterize the uncertainty in Casimir force measurements. 
Some of the sources of uncertainty are characteristic of ambient environments (water layers, drag, {\it etc.}), but many of the sources of error, such as interference artifacts and irregular transfer function from piezoelectric actuation, may appear in other environments as well. 
Comparing the measurements shown and characterized here to the force that should be observable by a thermal-noise limited measurement shows that the reduction of uncertainty could allow the Casimir force to be observed at separations up to 1.4 $\mu$m in this configuration. 
At separations in the 30 - 150 nm range, calibration and separation uncertainty dominate the error analysis, but under those considerations the data are consistent with the Lifshitz theory. 

Higher accuracy will assist the search for materials that can be used to electronically modulate the Casimir force, which could have many uses in future technologies, e.g. in next generation microelectromechanical systems \cite{Iannuzzi2005,Capasso2007,Capasso2011b}. 
	
We thank Mark Reitsma of Asylum Research for sending us light sources to test, thank Chen Gong for help with ellipsometry, and thank Dakang Ma, Sven de Man and Hedwig Eerkens for helpful discussions. 

We acknowledge funding from DARPA YFA grant number D18AP00060.
	
\bibliographystyle{iopart_num_edited}
\bibliography{Joes_Thesis}

\end{document}